\documentclass[twocolumn,floats,floatfix,showpacs,superscriptaddress,pre]{revtex4}
\usepackage{amsfonts,amsmath,amssymb,hyperref}
\usepackage{graphicx}
\usepackage{relsize}
\usepackage{bm}
\def\defi{{\buildrel \;def\; \over =}}

\newcommand{\mediaT}[1]{\left\langle #1 \right\rangle}
\newcommand{\mediaG}[1]{\left\{ #1 \right\}}
\newcommand{\mediaS}[1]{\left[ #1 \right]}

\begin{document}

\title{Fluctuations analysis in complex networks modeled by hidden variable models.
Necessity of a large cut-off in hidden-variable models}

\author{Massimo Ostilli}
\affiliation{Cooperative Association for Internet Data Analysis, University of California San Diego, CA, USA}

\begin{abstract}
It is becoming more and more clear that complex networks present remarkable large fluctuations. These fluctuations may manifest
differently according to the given model. In this paper we re-consider hidden variable
models which turn out to be more analytically treatable and for which we have recently shown clear evidence of non-self averaging;
the density of a motif being subject to possible uncontrollable fluctuations in the infinite size limit.
Here we provide full detailed calculations and we show that large fluctuations are only due to the node hidden variables variability while, in ensembles
where these are frozen, fluctuations are negligible in the thermodynamic limit, and equal the fluctuations of classical random graphs. 
%We then discuss how general are these results within the
%class of hidden-variables models, and what counterpart is expected to be found in other classes of models.
A special attention is paid to the choice of the cut-off: we show that in hidden-variable models, 
only a cut-off growing as $N^\lambda$ with $\lambda\geq 1$ can reproduce the scaling of a power-law degree distribution.
In turn, it is this large cut-off that generates non-self-averaging.
\end{abstract}

%\pacs{89.75.Hc, 89.75.Fb, 05.40.-a, 04.20.Gz}
\pacs{89.75.Hc, 89.75.Fb, 89.75.Kd, 05.40.-a}

\maketitle

\section{Introduction}
Complex networks are the output of certain non deterministic processes \cite{Bollobas,Barabasi,Dorog,NewmanBook}, therefore
they should be considered as intrinsically random objects. 
From this point of view statistical mechanics provides the best framework to analyze
and understand complex networks. 
Indeed, in the last years, several fundamental results on complex networks have been
derived by using ideas and techniques borrowed from statistical mechanics.
In particular, exact analysis has been done by applying concepts like micro-canonical and canonical ensembles and asymptotic techniques 
within the configuration model \cite{Samukhin}, and more general approaches and results have been derived via 
entropy maximization for both canonical (Shannon entropy) \cite{ERG}, 
and micro-canonical ensembles (Boltzmann entropy) \cite{Bianconi3}. 
However, not all aspects of statistical mechanics have been applied to complex networks so far.
A key concept in statistical mechanics is represented by the \textit{thermodynamic limit}:
a large system composed of $N$ elements at given concentration, is said to have
a thermodynamic limit when the relative fluctuations $R_\mathcal{O}$ of any physical observable $\mathcal{O}$, extensive in the system size $N$,
become negligible for large $N$
\begin{eqnarray}
\label{RI}
&& \lim_{N\to\infty}R_\mathcal{O} =0, \\
&& R_\mathcal{O}\defi \frac{\mediaT{\mathcal{O}^2}-\mediaT{\mathcal{O}}^2}{\mediaT{\mathcal{O}}^2}, \nonumber
\end{eqnarray}
where $\mediaT{\cdot}$ stands for sample-average, or expected value,
and the limit $N\to\infty$ is performed by keeping the concentration constant. 
In probability theory, Eq. (\ref{RI}) can be derived from the strong law of large numbers \cite{LyonsSLLN}: if 
$\mathcal{O}$ is a sum of identically distributed weakly dependent random variables, $\mathcal{O}=\sum_i x_i$,
the density $\sum_i x_i/N$ tends to the expected value of $x_i$, $\mediaT{x_i}$, with probability 1 and,
in particular, Eq. (\ref{RI}) applies. Vice-versa, if Eq. (\ref{RI}) is true, a law of large numbers applies,
(though not necessarily strong).
For example, on a regular lattice, an Ising model of $N$ spins coupled through ferromagnetic interactions at high enough
temperature has a thermodynamic limit. The existence of the thermodynamic limit in the Ising model
ensures that, regardless of any boundary conditions, the value of the density of the energy $E/N$ (or any extensive observable) ``tends'' always to a same number $u$.
Since the thermodynamic limit exists, we can ignore
the sample-to-sample fluctuations and, for a sufficiently large system, we can identify the actual value $E/N$ with its limit number $u$.
Notice that, rigorously speaking, $E/N$ is a random variable and the correct way to state the existence of the
thermodynamic limit is
\begin{eqnarray}
\label{TLI}
\left\{
\begin{array}{l}
\lim_{N\to\infty}\frac{E}{N}=u \quad \mathrm{a.s.}, \quad \mathrm{where} \\
u=\lim_{N\to\infty}\frac{\mediaT{E}}{N}.
\end{array}
\right.
\end{eqnarray}
%A counter example is the Ising model defined on a Cayley tree (not to be confused with the Bethe Lattice) \cite{BL}.
%It is well known that, for any $N$, the properties of such a system strongly depend on the
%value of the spins at the boundary so that, in particular, there is no thermodynamic limit. 

\begin{figure}[tbh]
{\includegraphics[height=4.5in]{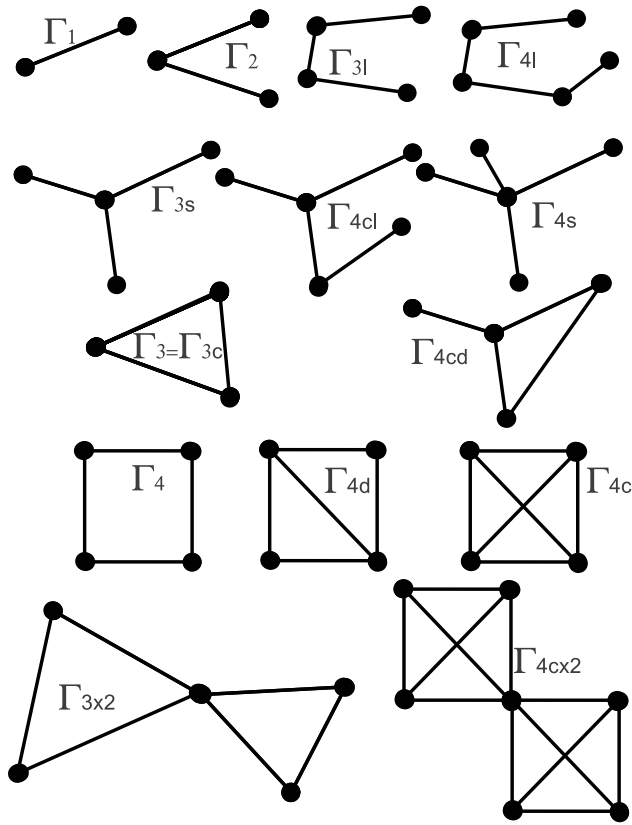}}
%{\includegraphics[height=4.5in]{figures/fig1_PRE.pdf}}
\caption{Examples of motifs. The labels in $\Gamma$ may specify the number of links of the motif;
for motifs made of $k$ fully connected nodes, $k$-cliques, we also use the symbol $\Gamma_{k\mathrm{c}}$; $d$ in $\Gamma_{\mathrm{4d}}$ stands for the presence of a diagonal;
$\Gamma_{kc\times 2}$ stands for two $k$-cliques sharing a common node; $\Gamma_{k\mathrm{s}}$ stands for a star-like motif with $k$ links;
$\Gamma_{k\mathrm{l}}$ stands for an open linear motif made of $k$ consecutive links.
\label{fig1}
}
\end{figure}

More in general, a system might not have a well defined thermodynamic limit for certain observables,
but have it for others. The latter are called self-averaging observables and are characterized by a ratio $R$ that goes to 0 when $N\to\infty$.
For example, in the Ising model, at low enough temperature the magnetization does not satisfy the analogous of Eq. (\ref{TLI}),
the magnetization being dependent on the two possible classes of boundary conditions (majority of the spins up or down, respectively). This kind of non self-averaging is trivial, but there are
more intriguing examples.
For instance, in mean-field models of spin-glasses, where the spins interact through couplings having
random signs, the free energy ``landscape'' is known to be rather ``rugged'', with many valleys
separated by high barriers whose height grows with the system size $N$.
From this picture, it is intuitively clear that 
the probability to find the system in a certain region of the phase space depends always on the system
sample, even in the limit of $N\to\infty$ \cite{Parisi,Derrida}. Nevertheless, the energy
of the system is self-averaging.
Upon applying the results of this paper, we will see that a similar picture holds also for complex networks.
Specifically, %we consider the class of hidden-variable models \cite{CaldaHidden,BogunaHidden,NewmanHidden,SatorrasHidden,BogunaHidden2}, and 
we analyze the behavior of the
relative fluctuations of the densities of the motifs $n_\Gamma\propto\sum_i k_\Gamma(i)/N$, where $\Gamma$ represent any motif (see Fig. \ref{fig1}),
and $k_\Gamma(i)$ is the generalized degree of node $i$ counting
how many motifs $\Gamma$ pass through it. We will see that, despite $n_\Gamma$ is an extensive observable (proportional to),
no strong law of large numbers applies. In order to make precise these statements, we need to introduce the concept of \textit{hidden-variable models}.

The processes that generate real complex networks are the result of certain complex dynamics governed by many internal (system) and external (environment) factors. 
A way to model these pseudo random processes consist in treating
them as hidden random processes characterized by some hidden variables whose distribution is supposed to be accessible.  
The family of models that best fit such a scheme is provided by the class of hidden-variable models
\cite{CaldaHidden,BogunaHidden,NewmanHidden,SatorrasHidden,BogunaHidden2}.
Depending on the context, the input in these models are either the expected degree of each node $\{h_i\}$, or their
probability density function (PDF) $\rho(h)$. In general, in a hidden-variable model, the resulting degree distribution
coincides with (is proportional to) $\rho(h)$ \cite{NewmanHidden} (though, as we will see, not all characteristics of $\rho(h)$ are reproduced if the cut-off is not
properly chosen).   
In view of applications, given a ``hidden'' process that generates real networks, it is crucially important to distinguish what kind of modeling we are pursuing at.
If we are interested in reproducing in simulations all possible graph realizations compatible with a specific real network, \textit{i.e.} a single output of the hidden random process, 
the input that we need to know are the expected degrees $\{h_i\}$, which must be treated as fixed parameters.
If we are instead interested in reproducing in simulations all possible network realizations as outputs of the whole hidden random process,
the expected degrees $\{h_i\}$ must be treated as random variables, and the input of the model is the PDF $\rho(h)$.
In \cite{SAEPL} we have shown that
in this latter case, if $\rho(h)\sim h^{-\gamma}$, and if we use a large cut-off $h_{\mathrm{max}}\geq N$, the fluctuations of the motif densities $n_\Gamma$ in general are not negligible.
In fact, given a motif $\Gamma$, there exists an interval $(\gamma_1,\gamma_2)$ where the relative fluctuations of $n_\Gamma$,
$R_{\Gamma}$, diverge in the thermodynamic limit,
where $\gamma_1\approx k_{\mathrm{min}}$ and $\gamma_2\approx 2 k_{\mathrm{max}}$, $k_{\mathrm{min}}$ and $k_{\mathrm{max}}$ being the smallest and the largest degree of $\Gamma$ (see Figs. \ref{fig2}-\ref{fig7}). 
By contrast, in classical random graphs $n_\Gamma$ is always self-averaging.

By using these results, we now understand the analogy with spin-glasses. Let us consider the space $\{n_{_\Gamma}\}$,
and the probability on the space $\{n_{_\Gamma}\}$, $\mathcal{P}\left(\{n_{_\Gamma}\}\right)$, where $\Gamma\in\mathcal{M}_N$,
and $\mathcal{M}_N$ is the set of all possible motifs compatible with the size $N$.
For fixed $N$, $\mathcal{P}\left(\{n_{_\Gamma}\}\right)$ provides the probability to find a network realization with
densities $\{n_{_\Gamma}\}$. In a classical random graph ensemble, for $N$ large enough we have ($\delta(x,y)$ stands for Kronecker' s delta function)
\begin{eqnarray}
\label{Png}
\mathcal{P}^{\mathrm{Classical}}\left(\{n_{_\Gamma}\}\right) \sim \prod_{\Gamma\in\mathcal{M}_N}\delta(n_\Gamma,\mediaT{n_\Gamma}),
\end{eqnarray} 
whereas, due to the non-self-averaging behavior for models characterized by power laws, we have
\begin{eqnarray}
\label{Png1}
\mathcal{P}^{\mathrm{Power-law}}\left(\{n_{_\Gamma}\}\right)\neq \prod_{\Gamma\in\mathcal{M}_N}\delta(n_\Gamma,\mediaT{n_\Gamma}).
\end{eqnarray} 
Eq. (\ref{Png1}) is the complex network analog of the aforementioned rugged spin-glass picture: the probability to find the network
in a given configuration $\{n_{_\Gamma}\}$ depends always on the specific sample, even if $N\to\infty$.
In this paper we show that the non-self-averaging scenario of Eq. (\ref{Png1}) is due only to the variability of the expected degrees of the nodes $\{h_i\}$.
In fact, we will show that, when the sequence $\{h_i\}$ is supposed to be fixed, \textit{i.e.},
when we want to reproduce all the possible graphs compatible with the fixed sequence $\{h_i\}$, 
the fluctuations are small, and the standard law of large numbers recovered for any observable: 
$R_{\mathcal{O}}\sim 1/\sqrt{N}$. As a consequence, when the sequence $\{h_i\}$ is fixed, Eq. (\ref{Png}) applies like
in the classical random graph~\footnote{
This property is consistent with the observed fact \cite{SatorrasHidden} that the conditional
probability for a node $i$ to have degree $k_i$, conditioned on the expected degree $h_i$,
is a Poissonian distribution, whose fluctuations, as known, are standard.
}. 
We will see however also that, as a result of the non-self-averaging scenario, even if we are dealing with a case of fixed expected degrees, 
the fact that our knowledge on the expected degree sequence $\{h_i\}$ is affected by unavoidable finite errors,
will result in a systematic bias between the simulations and the real network.

Quite interestingly, we will also see that we recover the limit (\ref{Png}) both when $\gamma\to\infty$,
and when $\gamma\to2$. Whereas the former is intuitively expected since $\gamma\to\infty$ corresponds, roughly speaking, to a an exponential
decay $P(k)\sim \exp(-bk)$, \textit{i.e.}, it is a classical random graph limit,
the latter is due to the fact that entropy of power-law random graphs goes to 0 when $\gamma\to2^+$ \cite{Bianconi3,Genio}.

The ultimate mathematical reason for the non-self-averaging behavior lies in the necessity to choose a large cut-off in the hidden-variable models.
In fact, as we will see in Sec. II, if a target network is characterized by a power-law degree distribution $P(k)\sim k^{-\gamma}$,
a hidden-variable model with PDF $\rho(h)\sim h^{-\gamma}$ reproduces the correct scaling behavior of $P(k)$ only if we choose a cut-off
$h_{\mathrm{max}}=N^\lambda$ with $\lambda\geq 1$. To the best of our knowledge, this important fact has remained unknown so far.
In Sec. II we will make it clear that, for any $\gamma$, choosing $\lambda<1$ leads to an incorrect scaling of the moments of the sampled degree in the model. 
By contrast, in literature, the so called natural cut-off $h_{\mathrm{max}}=N^{1/(\gamma-1)}$, or the structural cut-off $h_{\mathrm{max}}=N^{1/2}$, were often used
for $\gamma\leq 3$, or $\gamma> 3$, respectively. In fact, in \cite{Bianconi1,Bianconi2,BogunaC} fluctuations of a few motifs have been
analyzed but, as a consequence of the small cut-off, $R_\Gamma$ remained finite.

If networks characterized by power law distributions are non self-averaging, network configurations, like in spin-glass models,
are intrinsically unpredictable. Furthermore, unless $\gamma$ is close to 2, or very large, the broad distribution (\ref{Png1}) makes a
power-law network effectively unstable to small perturbations. This in particular reflects on the stability/instability of communities \cite{Santo},
and more in general to the stability/instability of the motifs on which the functionality of the network largely depends \cite{Milo,BaraProc,BaraBio}.

The paper is organized as follows.
In Sec. II we review the definition of hidden-variable models and show the necessity 
to choose a cut-off that scales as $N^\lambda$, with $\lambda\geq 1$. 
%We will see in fact
%that a cut-off with $\lambda<1$ cannot reproduce correctly the power-law target distribution. 
In Sec. III we distinguish between node-hidden variables and link-hidden variables and define
three different ways of sampling according to the order we sample the two sets of hidden variables.
In Sec. IV for each ensemble we introduce the proper relative fluctuation ratio.
In Sec. V we collect and state the main result.
Finally in Sec. VI we provide detailed proofs of the Eqs. listed in Sec. V, as well as the analysis
of $R$ for specific motifs, whose plots are reported in Figs. (\ref{fig2}-\ref{fig8}).
The reader not interested in detailed diagrammatic calculations may skip the reading of Sec. V.
At the end of the paper some conclusions are finally drawn.

Part of the results reported in this paper were shortly presented in the Letter \cite{SAEPL}.
In the present paper, besides the novel content, we provide full analysis of several aspects
that were only shortly mentioned in \cite{SAEPL}.

\section{Hidden-variable models. 
Necessity of a cut-off growing as $N^\lambda$ with $\lambda\geq 1$}
\begin{figure}[tbh]
{\includegraphics[height=2.3in]{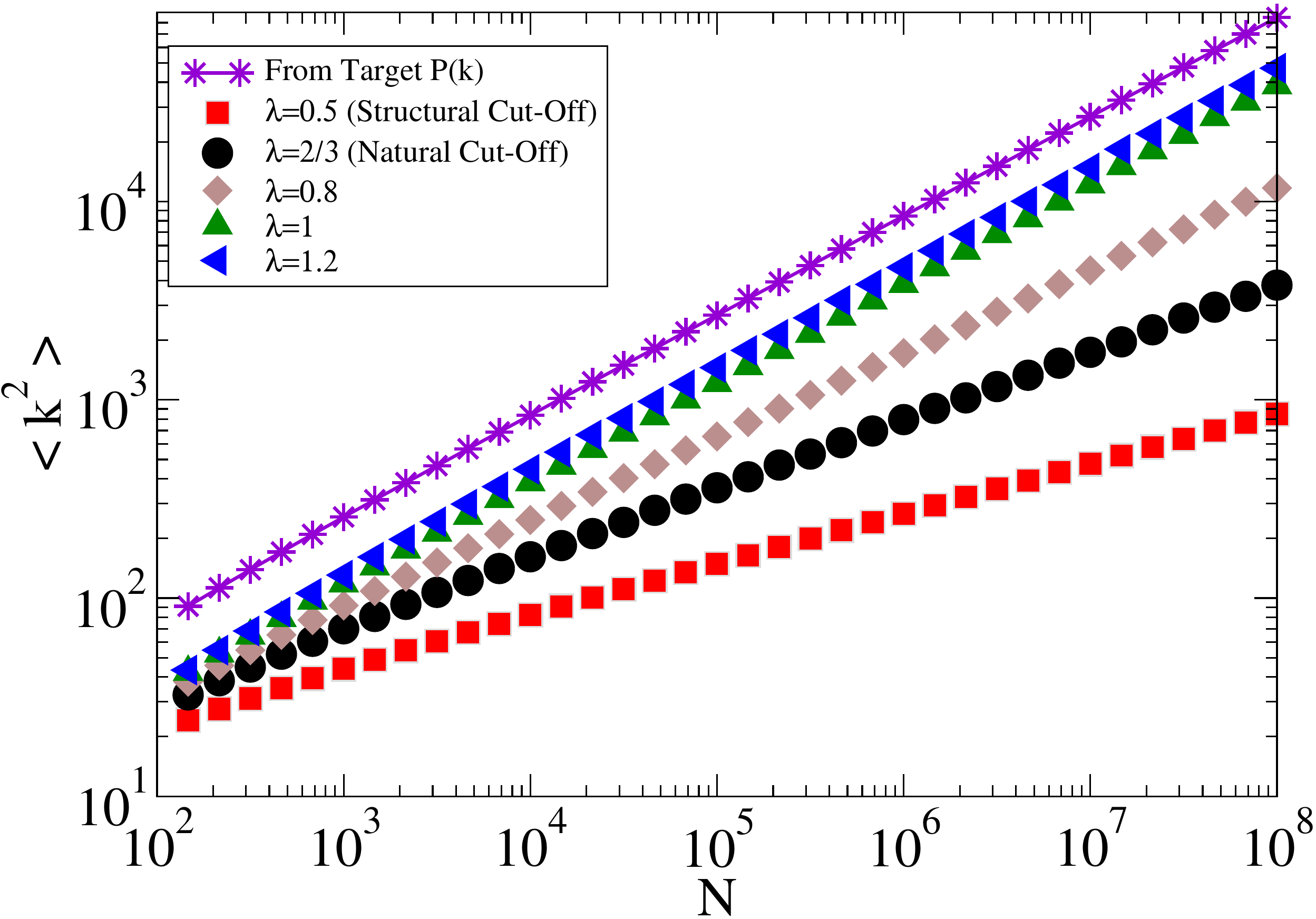}}
%{\includegraphics[height=2.4in]{figures/figkk_PRE.pdf}}
\caption{Behavior of $\mediaT{k^2}$ \textit{vs} the system size $N$ for $\gamma=2.5$. The upper plot corresponds to the target degree distribution which is a
pure power law $P(k)=ak^{-\gamma}$,
whereas the other plots correspond to the hidden variable model (\ref{CM})-(\ref{rho}) with different choices of the cut-off $h_{\mathrm{max}}=N^\lambda$: 
$\lambda=0.5$ (corresponding to the ``structural cur-off''), $\lambda=1/(\gamma-1)=2/3$ (corresponding to the ``natural cut-off''), $\lambda=0.8$, $\lambda=1$, and $\lambda=1.2$. For higher values of $\lambda$,
the plots saturate to a curve that, on this scale, is indistinguishable from the case $\lambda=1.2$. 
The plots of the hidden-variable model have been calculated by numerical evaluation of the involved integrals:
$\mediaT{k^2}\simeq N^2\int_{h_{\mathrm{min}}}^{h_{\mathrm{max}}}dh_1dh_2dh_3\rho(h_1)\rho(h_2)\rho(h_3)p(h_1,h_2)p(h_1,h_3)$ (see below for a more detailed analysis
of these techniques).
\label{figkk_PRE}
}
\end{figure}

\begin{figure}[tbh]
{\includegraphics[height=2.3in]{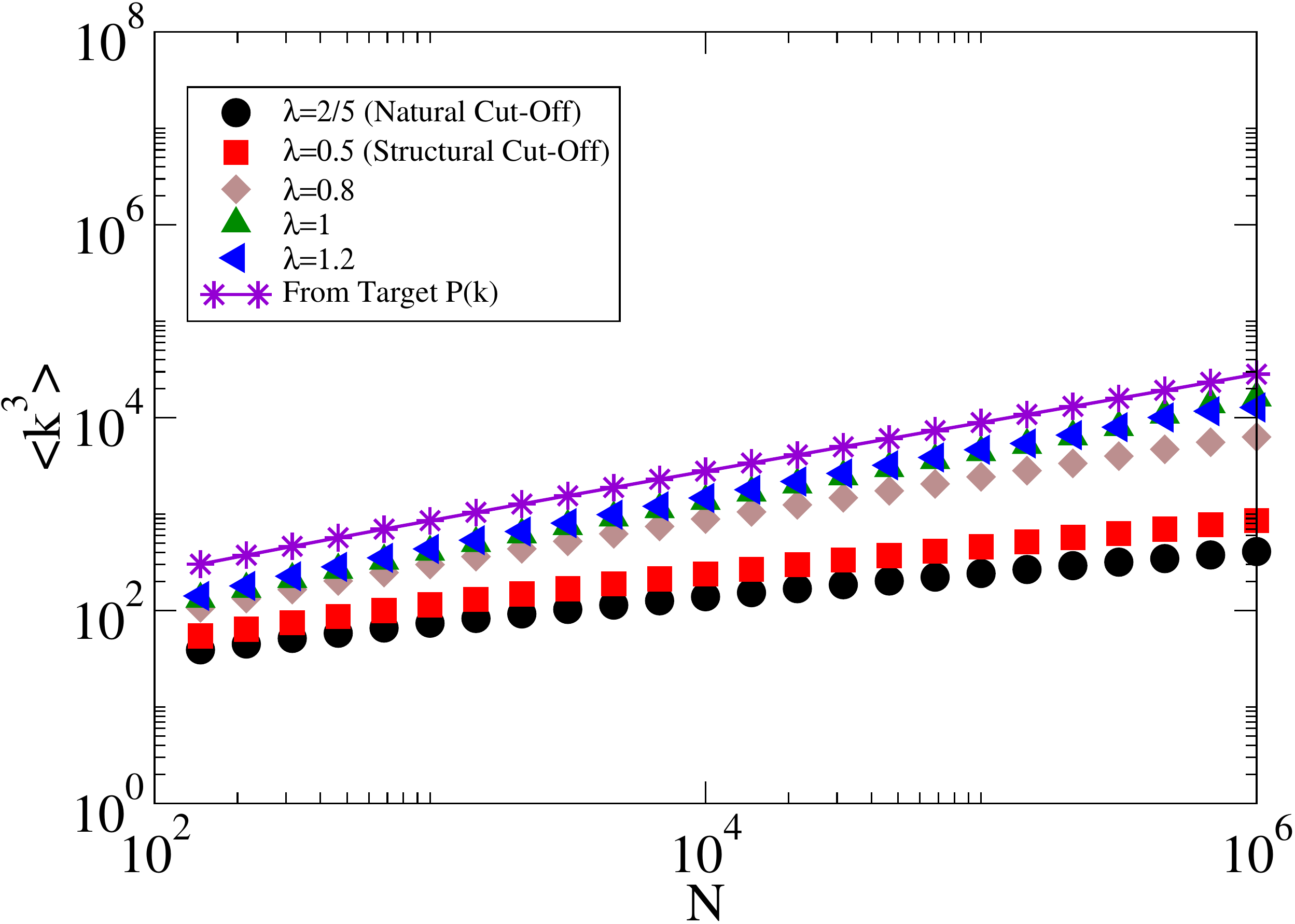}}
%{\includegraphics[height=2.4in]{figures/figkkk_PRE.pdf}}
\caption{Behavior of $\mediaT{k^3}$ \textit{vs} the system size $N$ for $\gamma=3.5$. The upper plot corresponds to the target degree distribution which is a
pure power law $P(k)=ak^{-\gamma}$,
whereas the other plots correspond to the hidden variable model (\ref{CM})-(\ref{rho}) with different choices of the cut-off $h_{\mathrm{max}}=N^\lambda$: 
$\lambda=1/(\gamma-1)=2/5$ (corresponding to the ``natural cut-off''), $\lambda=0.5$ (corresponding to the ``structural cur-off''), $\lambda=0.8$, $\lambda=1$, and $\lambda=1.2$. For higher values of $\lambda$,
the plots saturate to a curve that, on this scale, is indistinguishable from the case $\lambda=1.2$. 
The plots of the hidden-variable model have been calculated by numerical evaluation of the involved integrals:
$\mediaT{k^3}\simeq N^3\int_{h_{\mathrm{min}}}^{h_{\mathrm{max}}}dh_1dh_2dh_3dh_4\rho(h_1)\rho(h_2)\rho(h_3)\rho(h_4)p(h_1,h_2)$ $\times p(h_1,h_3)p(h_1,h_4)$ (see below for a more detailed analysis
of these techniques).
\label{figkkk_PRE}
}
\end{figure}

Hidden-variable models are ``soft'' models where, unlike ``hard'' models, graph constrains
are satisfied only on average. Hidden-variable models
have the remarkable advantage to be analytically treatable. In fact, in this paper
it will be evident that the average of almost any observable can be reduced to suitable
integrals. We now review briefly the definition of hidden-variable models. 
Given $N$ nodes, hidden variable models are defined in the following way:
\textit{i)}
to each node we associate a hidden variable $h$ drawn from
a given probability density function (PDF) $\rho(h)$; 
\textit{ii)} between any pair of nodes, we assign, or not assign, a link, according
to a given probability $p(h,h')$, where $h$ and $h'$ are the hidden variables associated to the two nodes.
The probability $p(h,h')$ can be any function of the $h$'s, the only requirement being that $0 \leq p(h,h')\leq 1$.
The hidden variables can be real numbers or also vectors. Many networks can be embedded in a hidden-variable scheme.
For example, in random geometric graphs the $h$' s are vectors representing the positions of the nodes,
and the $p(h,h')$'s are defined in terms of geometric rules (non deterministic if $0<p(h,h')<1$, or deterministic if $p(h,h')$ takes only the values 0 or 1).
Particular attention has been paid to the ``configuration model''. In this case
$p(h,h')$ has the following form (or similar generalizations)~
\begin{eqnarray}
\label{CM}
&& p(h,h')= \left(1+\frac{k_s^2}{hh'}\right)^{-1}, \quad
k_s=\sqrt{N\bar{k}},
\end{eqnarray}
where $\bar{k}$ is the wanted average degree, for large $N$. In general, the actual degree $k$ of the nodes of the network realized with the above scheme
are distributed according to $\rho$ with actual average degree equal to $\bar{k}$.
In particular, if we choose the following PDF having support in $[h_{\mathrm{min}},h_{\mathrm{max}}]$
\begin{eqnarray}
\label{rho}
\rho(h)=a~h^{-\gamma}, \quad h_{\mathrm{max}} \geq h \geq h_{\mathrm{min}}>0,
\end{eqnarray}
with $\gamma>2$,
the degree-distribution $P(k)$ of the resulting network will be a power law with exponent $\gamma$ and, 
the normalization constant $a$, and the so called structural cut-off $k_s$ are
\begin{eqnarray}
\label{a}
a=\frac{\gamma-1}{\bar{k}_{\mathrm{min}}^{1-\gamma}-N^{1-\gamma}},
\end{eqnarray}
\begin{eqnarray}
k_s=\sqrt{N\left(\frac{\gamma-1}{\gamma-2}\right)\frac{\bar{k}_{\mathrm{min}}^{2-\gamma}-N^{2-\gamma}}{\bar{k}_{\mathrm{min}}^{1-\gamma}-N^{1-\gamma}}}.
\end{eqnarray}
%which for $N$ sufficiently large become
%\begin{eqnarray}
%\label{struc}
%&& a=\frac{\gamma-1}{h_{\mathrm{min}}^{1-\gamma}}, \\
%&& k_s=\sqrt{Nh_{\mathrm{min}}\frac{\gamma-1}{\gamma-2}}.
%\end{eqnarray}
When $h_{\mathrm{max}}\ll k_s$, correlations of the generated network are negligible, and $p(h,h')\simeq hh'/k_s^2$,
while for $h_{\mathrm{max}}\gg k_s$ correlations can be important. 
The choice of the cut-off $h_{\mathrm{max}}$ may be in principle arbitrary. 
However, most of real-world networks show that the maximal degree scales according to the so called natural cut-off: $k_{\mathrm{max}}\sim h_{\mathrm{nc}}=N^{1/(\gamma-1)}$. 
A consequence of this observation was that in several models of complex networks it was assumed the choice $h_{\mathrm{max}}=h_{\mathrm{nc}}$,
justified as empirical. We think however that such an approach is wrong: the fact that in most of the real-world networks  
$k_{\mathrm{max}}\sim h_{\mathrm{nc}}$ is due to a probabilistic effect, is not due to a rigid upper bound $k\leq h_{\mathrm{nc}}$. 
In fact, by using order-statistics one finds
that by drawing $N$ degree values from a power law with exponent $\gamma$,
the highest degree in average scales just as $\mediaT{k_{\mathrm{max}}}\sim N^{1/(\gamma-1)}$ \cite{DorogCut,BogunaCut}.
More precisely, it is possible to prove that the PDF for the rescaled random variable $k_{\mathrm{max}}/N^{1/(\gamma-1)}$ 
is also a power law with exponent $\gamma$ \cite{Remco}. 
Power law distributions always lead to important fluctuations. 
It is then clear that empirical observations of $k_{\mathrm{max}}$ must be taken with care: 
$k_{\mathrm{max}}$ is not a self-averaging variable and samples in which $k_{\mathrm{max}}\gg N^{1/(\gamma-1)}$,
even if extremely rare, do exist and, as we shall see, have dramatic effects on the fluctuations of motifs. 
We stress that here we follow a null-model approach. We do not claim that the degree of all real networks must have a cut-off scaling
with $N$; there might be of course many other possible scalings whose value depend on the details of the system (related to physical, biological, or economical constraints). 
However, if the only information that we have from a given real network of size $N$ is that \textit{i)} the degree obeys a power-law distribution with exponent $\gamma$, 
and \textit{ii)}  highest degrees scale in average as $N^{1/(\gamma-1)}$,    
forcing the model to have a specific cut-off other than $N$ would introduce a bias. In fact, order statistics tells us that a lower cut-off would
produce $\mediaT{k_{\mathrm{max}}}\ll N^{1/(\gamma-1)}$.
This observation leads us to choose $h_{\mathrm{max}}\sim N$ for the hidden-variable scheme (\ref{CM})-(\ref{rho}). 
More precisely, if we consider as target degree distribution a power-law $P(k)\propto k^{-\gamma}$ having finite support $k\leq N$,  
in order to reproduce its characteristics from the hidden-variable model
we need to use a cut-off $h_{\mathrm{max}}=\mathrm{O}(N^\lambda)$ with $\lambda\geq 1$. In fact, any other choice implies a difference
in the scaling of the moments $\mediaT{k^n}$ between the hidden variable model (\ref{CM})-(\ref{rho}) and the target distribution $P(k)$.
Fig. \ref{figkk_PRE} shows this property for $\mediaT{k^2}$ and $\gamma<3$, whereas Fig. \ref{figkkk_PRE} show a case for $\mediaT{k^3}$ and $4>\gamma>3$.  
Similar plots hold for higher moments. In general, for any $\gamma$ there is exists a minimal exponent $n_c=\gamma-1$ such that,
to reproduce correctly the behavior of $\mediaT{k^n}$ with $n>n_c$ we need $\lambda\geq 1$.
Notice that larger cut-offs do not change the moments $\mediaT{k^n}$, but result in larger fluctuations
of the motif densities $n_\Gamma$. 
In conclusion, the minimal cut-off of the model (\ref{CM})-(\ref{rho})
able to reproduce the correct scaling of all the moments of the target degree distribution $P(k)\sim k^{-\gamma}$ is just $h_{\mathrm{max}}=\mathrm{O}(N^\lambda)$ with $\lambda=1$. 
In this paper we set therefore $h_{\mathrm{max}}=N$. We stress  
that with this choice highest degrees will be still order $N^{1/(\gamma-1)}$, but only on average.  

\section{Node-hidden-variables and link-hidden-variables. Three different ways of sampling}
An attentive observation of the hidden-variable scheme shows us that, actually, for each graph realization,
besides the set of the $N$ hidden variables $h_1,\ldots,h_N$, that from now on we shall call the node-hidden-variables,
we need also the set of the $N(N-1)/2$ link-hidden-variables $q_{1,2},\ldots,q_{N-1,N}$, where each $q$ is drawn independently
from the uniform distribution on the interval $[0,1]$. In fact, for each couple of nodes $i$ and $j$, with associated
the node-hidden-variables $h_i$ and $h_j$, respectively, we need to sample the corresponding probability $p(h_i,h_j)$
in order to establish the presence or not of a link between the nodes $i$ and $j$. To this aim we draw a random number
$q_{i,j}$ uniformly distributed on the interval $[0,1]$, and the nodes $i$ and $j$ are connected through a link
if $q_{i,j}<p(h_i,h_j)$, while they remain not connected if $q_{i,j}>p(h_i,h_j)$.
All the possible graphs are in correspondence with all the possible combinations of these two sets of hidden-variables.
However, due to the full independence of the two sets, both within the individual sets and with each other, 
we are free to sample the graph realizations in three different ways.

\subsection{Symmetric sampling (the traditional way)}
In literature, the hidden-variables model is becoming more and more popular, enriched by several generalizations and uses. 
However, the usual ways these ensembles are sampled consists in keeping track of the sole node-hidden-variables set.
This way the two sets of hidden-variables are simultaneously, or symmetrically, drawn.
In \cite{SAEPL} we have dealt with this case. In this paper we will use the upper script $^{(S)}$ to indicate
that we are referring to the symmetric case, and we keep using the symbol $\mediaT{\cdot}$ to indicate
the ensemble averages obtained in this symmetric way. We shall soon see that we can call the $\mediaT{\cdot}$ averages
also ``full-averages''.

\subsection{Sampling by first freezing the $h$'s}
We are free to sample the ensemble in the following alternative way.
For any extraction of the node-hidden-variables, $\{h_i\}$ (where $i=1,\ldots N$), we draw many times the set of the link-hidden-variables $\{q_{(i,j)}\}$ 
(where $(i,j)=(1,2),\ldots, (N-1,N)$),
and we calculate the corresponding averages with frozen $\{h_i\}$, $\mediaT{\cdot}_{|\bm{h}}$.
We then proceed with many extractions of the $h$'s and at the end we calculate the full averages.
According to the formalism used in \cite{SAEPL}, we will indicate by $\mediaS{\cdot}$ the averages with respect to the $h$'s,
\textit{i.e.}, if $f(\cdot)$ is any function of the node-hidden-variables, we define
\begin{eqnarray}
\label{aveh}
\left[f\right]=\int \prod_{i=1}^N dh_i\rho(h_i)~f(\cdot).
\end{eqnarray}
In particular, we have
\begin{eqnarray}
\label{avea}
\mediaT{a_{i,j}}=\left[p(h,h')\right]=\mathrm{O}\left(\frac{1}{N}\right),
\end{eqnarray}
where in the last term we have used the fact the network is sparse.

\subsection{Sampling by first freezing the $q$'s}
We are free to sample the ensemble in this further alternative way.
For any extraction of the link-hidden-variables, $\{q_{i,j}\}$, we draw many times the set of the nodes-hidden-variables $\{h_i\}$,
and we calculate the corresponding averages with frozen $\{q_{i,j}\}$, $\mediaT{\cdot}_{|\bm{q}}$.
We then proceed with many extractions of the $q$'s and at the end we calculate the full averages.
%We will indicate by $\overline{\cdot}$ the averages with respect to the $q$'s.
We will indicate by $\mediaG{\cdot}$ the averages with respect to the $q$'s,
\textit{i.e.}, if $g(\cdot)$ is any function of the link-hidden-variables, we define
\begin{eqnarray}
\label{aveq}
\mediaG{g}=\int_{\bm{\left[0,1\right]}} \prod_{i<j} dq_{i,j}~g(\cdot),
\end{eqnarray}
where $\bm{\left[0,1\right]}$ is a short notation for the $N(N-1)/2$ hypercube $\bm{\left[0,1\right]}=\left[0,1\right]\times\cdots\times \left[0,1\right]$.

\subsection{Equivalence and differences} 
Due to the full independence of the two sets of random variables, the $h$'s and the $q$'s,
it is immediate to see that, if $\mathcal{O}$ is any graph observable
\begin{eqnarray}
\label{Equi}
\mediaT{\mathcal{O}}=\mediaS{\mediaT{\mathcal{O}}_{|\bm{h}}}=\mediaG{\mediaT{\mathcal{O}}_{|\bm{q}}}.
\end{eqnarray}
However, the same equivalence does not apply for higher moments.
In particular, if $\mathop{O_1}$ and $\mathop{O_2}$ are any two observables, we have
\begin{eqnarray}
\label{NEqui}
\mediaT{\mathcal{O}_1\mathcal{O}_2}\neq \mediaS{\mediaT{\mathcal{O}_1}_{|\bm{h}}\mediaT{\mathcal{O}_2}_{|\bm{h}}}\neq \mediaG{\mediaT{\mathcal{O}_1}_{|\bm{q}}\mediaT{\mathcal{O}_2}_{|\bm{q}}}.
\end{eqnarray}
In the following we will refer to these three ensembles as symmetric (S), $h$-quenched (or A), and $q$-quenched (or B).

\section{Fluctuations of Motifs}
Given the parameters $N$, $h_{\mathrm{min}}$, $\bar{k}$, and $\gamma$,
the hidden-variables scheme produces an ensemble of networks which, in terms of a few characteristics, like statistics
of the degree and motifs, are in part representative of many real-world networks with those given parameters. 

Following the formalism already used in \cite{SAEPL},
we will indicate by $n_\mathsmaller{\Gamma}$ the density of the motif $\Gamma$ in a network realization. 
As is known \cite{BogunaHidden,BogunaC}, for $\gamma>2$, the hidden variable model defined through Eqs. (\ref{CM})-(\ref{rho}) leads to a small clustering coefficient
$C=\mediaT{n_\mathsmaller{\Gamma_3}}/\mediaT{3n_\mathsmaller{\Gamma_2}}$.
For example, for $\gamma\gg 3$ we have $\mediaT{n_\mathsmaller{\Gamma_2}}=\mathrm{O}(1)$, while
$\mediaT{n_\mathsmaller{\Gamma_3}}=\mathrm{O}(1/N)$. 
%(solution for any $\gamma$ can be found in \cite{BogunaC}).
%whereas the density of triples $n_\mathsmaller{\Gamma_2}$ remains finite, 
%the density of triangles $n_\mathsmaller{\Gamma_3}$ decays as $1/N^\alpha$,
%with $\alpha=3$ for $\gamma$ sufficiently large. 
More in general, the more the motif is clustered, the smaller is its density.
Yet, for finite $N$, and for any motif $\Gamma$, clustered or not, 
by tuning the parameters $h_{\mathrm{min}}$, $\bar{k}$ and $\gamma$, one can set, within some freedom, 
a desired value of $\mediaT{n_\mathsmaller{\Gamma}}$.
%Yet, for finite $N$, and for any motif $\Gamma$, 
%by tuning the parameters $\bar{k}$, and $\gamma$, one can set, within some freedom, 
%a desired value of $\mediaT{n_\mathsmaller{\Gamma}}=A_\mathsmaller{\Gamma}/N^{\alpha_\Gamma}$, 
%where $n_\mathsmaller{\Gamma}$ is the density of the motif $\Gamma$ in the graph, $\alpha_\Gamma$ a suitable exponent ($\alpha_{\Gamma_1}=0$)
%and, for $\gamma>2$, $A_\mathsmaller{\Gamma}$ smoothly depends on $N$.
The sample-to-sample fluctuations of $n_\mathsmaller{\Gamma}$, however, 
can be unexpectedly large. Our task is to analyze these fluctuations in each possible ensemble, with or without quenched variables.
The fluctuations of $n_\mathsmaller{\Gamma}$ must be compared with the corresponding average of $n_\mathsmaller{\Gamma}$,
therefore, for each ensemble, we are going to analyze the following standard ratios
\begin{eqnarray}
\label{RS}
R^{(\mathrm{S})}_\mathsmaller{\Gamma}=\frac{\mediaT{{n_\mathsmaller{\Gamma}}^2}-\mediaT{n_\mathsmaller{\Gamma}}^2}{\mediaT{n_\mathsmaller{\Gamma}}^2},
\end{eqnarray}
\begin{eqnarray}
\label{RA}
R^{(\mathrm{A})}_\mathsmaller{\Gamma}=\frac{\mediaS{\mediaT{\mathop{n_\mathsmaller{\Gamma}}}^2_{|\bm{h}}}-\mediaS{\mediaT{\mathop{n_\mathsmaller{\Gamma}}}_{|\bm{h}}}^2}
{\mediaS{\mediaT{\mathop{n_\mathsmaller{\Gamma}}}_{|\bm{h}}}^2},
\end{eqnarray}
\begin{eqnarray}
\label{RB}
R^{(\mathrm{B})}_\mathsmaller{\Gamma}=\frac{\mediaG{\mediaT{\mathop{n_\mathsmaller{\Gamma}}}^2_{|\bm{q}}}-\mediaG{\mediaT{\mathop{n_\mathsmaller{\Gamma}}}_{|\bm{q}}}^2}
{\mediaG{\mediaT{\mathop{n_\mathsmaller{\Gamma}}}_{|\bm{q}}}^2}.
\end{eqnarray}
It will be also interesting to consider the local fluctuations for frozen hidden-variables, via the following standard quenched (Q) ratios
\begin{eqnarray}
\label{Rh}
R^{(\mathrm{Q})}_{\mathsmaller{\Gamma}|\bm{h}}=\frac{\mediaT{\mathop{n_\mathsmaller{\Gamma}}^2}_{|\bm{h}}-\mediaT{\mathop{n_\mathsmaller{\Gamma}}}_{|\bm{h}}^2}
{\mediaT{\mathop{n_\mathsmaller{\Gamma}}}_{|\bm{h}}^2},
\end{eqnarray}
\begin{eqnarray}
\label{Rq}
R^{(\mathrm{Q})}_{\mathsmaller{\Gamma}|\bm{q}}=\frac{\mediaT{\mathop{n_\mathsmaller{\Gamma}}^2}_{|\bm{q}}-\mediaT{\mathop{n_\mathsmaller{\Gamma}}}_{|\bm{q}}^2}
{\mediaT{\mathop{n_\mathsmaller{\Gamma}}}_{|\bm{q}}^2},
\end{eqnarray}
as well as their averages $\mediaS{R^{(\mathrm{Q})}_{\mathsmaller{\Gamma}|\bm{h}}}$ and $\mediaG{R^{(\mathrm{Q})}_{\mathsmaller{\Gamma}|\bm{q}}}$.

In \cite{SAEPL} we have shown that 
$R^{(\mathrm{S})}_\mathsmaller{\Gamma}$ strongly depends on $\Gamma$, $N$ and $\gamma$.
We will see that different scenario may apply for each different ensemble analyzed through $R^{(\mathrm{X})}_\mathsmaller{\Gamma}$,
where $X$ can be $X=S$ (symmetric), $X=A$ (by first freezing the $h$'s and then averaging over these latter), $X=B$ (by first freezing the $q$'s and then averaging over these latter),  
or $X=Q$ (for frozen $h$'s, or frozen $q$'s), according to Eqs. (\ref{RS}-\ref{Rq}), respectively.
In general, when $R^{(\mathrm{X})}_\mathsmaller{\Gamma}\to 0$ for $N\to\infty$, the ensemble X is said to be self-averaging with respect to the motif density $n_\mathsmaller{\Gamma}$.
In practical terms, when this occurs, even one single sample is enough to get, by simulations, an accurate estimation of the ensemble-average of $n_\mathsmaller{\Gamma}$, 
provided $N$ is large enough. The behavior of $R^{(\mathrm{X})}_\mathsmaller{\Gamma}$ with respect to the network size $N$ is therefore of crucial importance:
if the network is not self-averaging with respect to some motif $\Gamma$, the number of samples necessary to get 
a good estimation of the ensemble-average of $n_\mathsmaller{\Gamma}$ in simulations, will have to grow with $N$.
For the same reason, it will be hard to generate only those samples whose density is close to a target value, 
and a kind of hard searching problem emerges. This latter aspect is in fact connected 
with spin-glass and NP-complete problems; in Sec. VI we will see in fact that $R^{(\mathrm{X})}_\mathsmaller{\Gamma}$
can be read as a susceptibility of a homogeneous system.

\section{Main result}

\begin{widetext}

\begin{figure}[tbh]
{\includegraphics[height=3.5in]{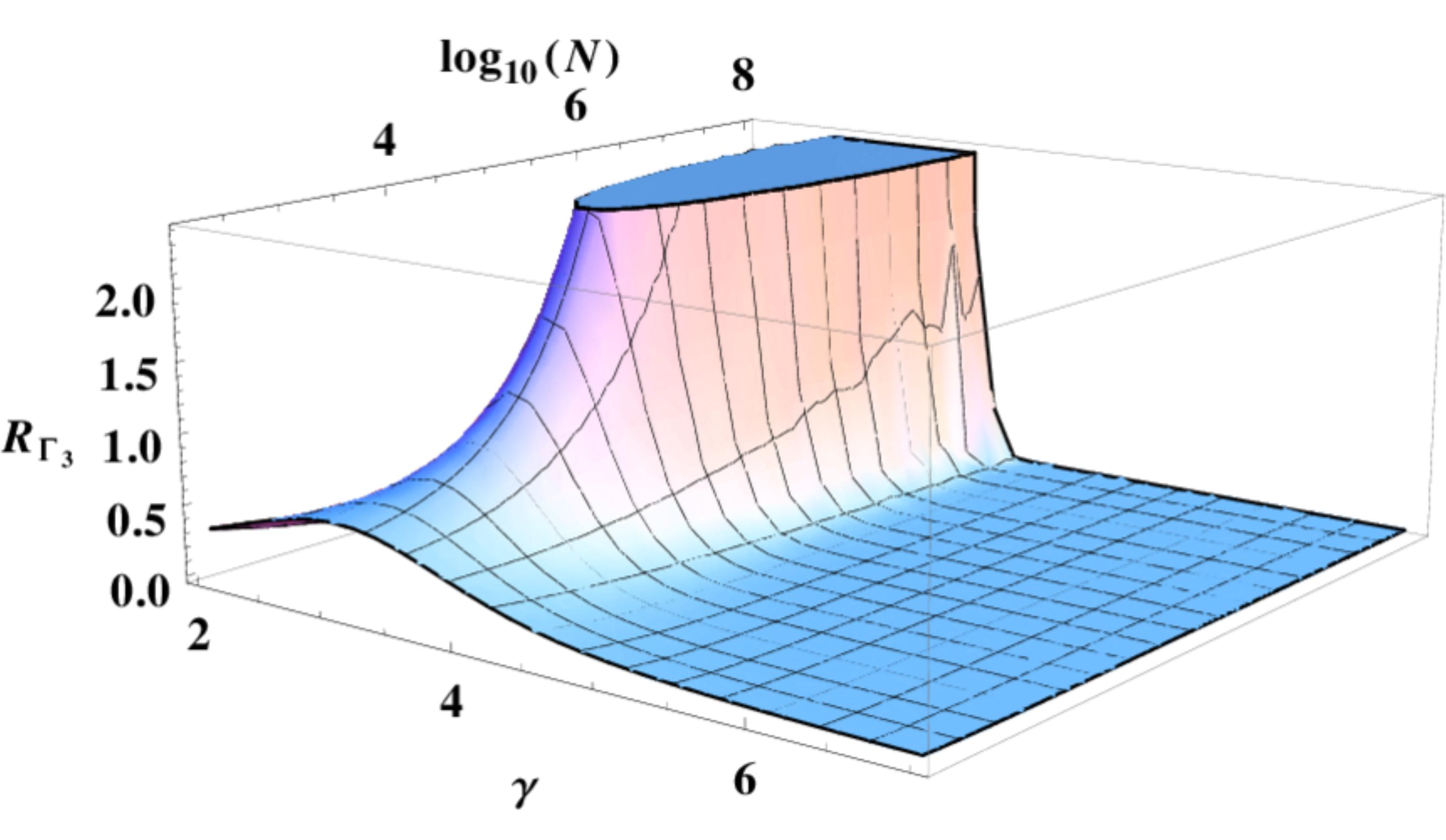}}
%{\includegraphics[height=3.5in]{figures/fig2_PRE.pdf}}
\caption{Behavior of $R^{(A)}_{\Gamma_3}$, as a function of $\gamma$ and $N$,
obtained from numerical integration of Eq. (\ref{RTA}) by using $10^7$ points per integral. 
\label{fig2}
}
\end{figure}
\begin{figure}[tbh]
{\includegraphics[height=3.5in]{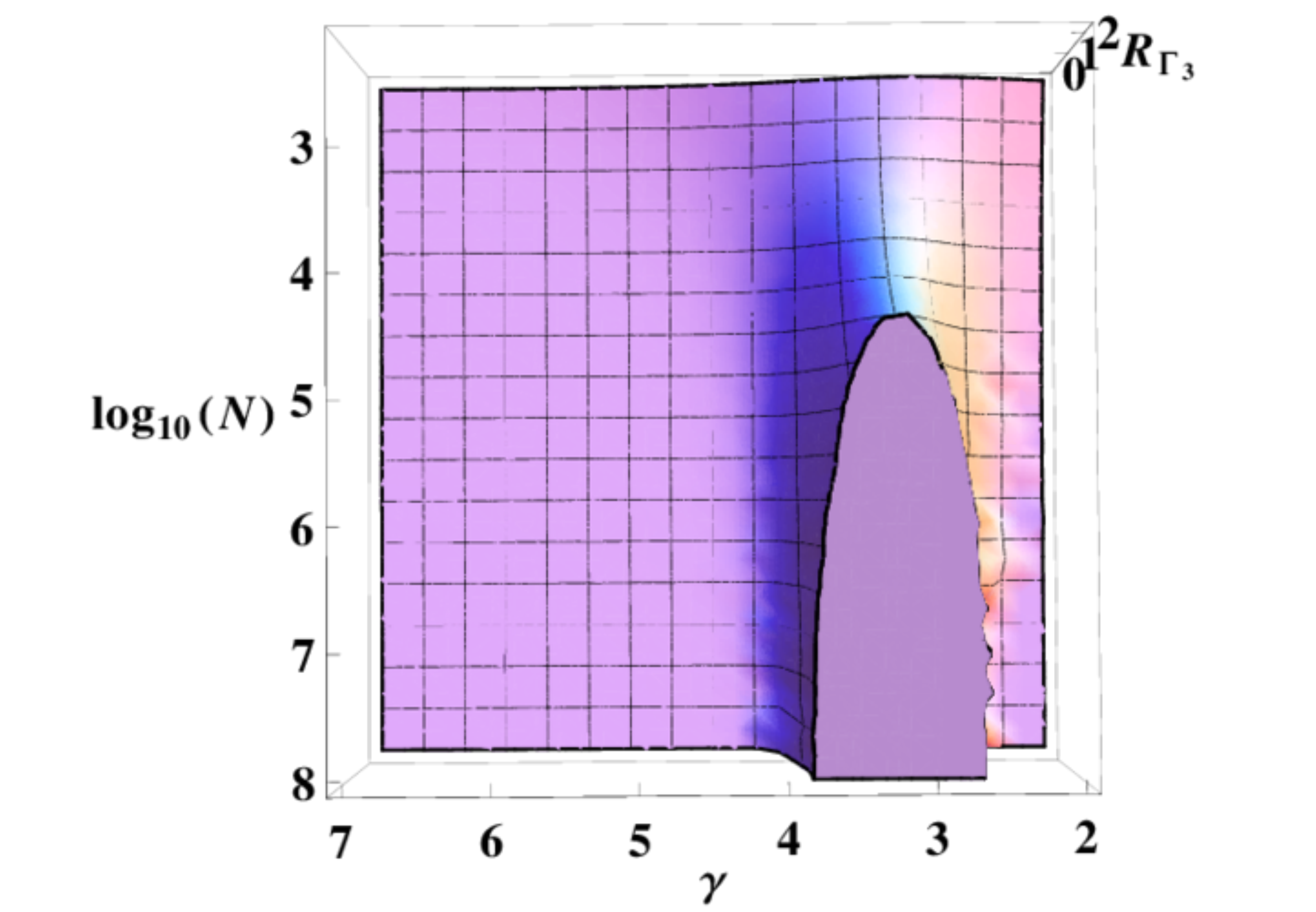}}
%{\includegraphics[height=3.5in]{figures/fig3_PRE.pdf}}
\caption{Top view of Fig. \ref{fig2}.
\label{fig3}
}
\end{figure}
\begin{figure}[tbh]
{\includegraphics[height=3.5in]{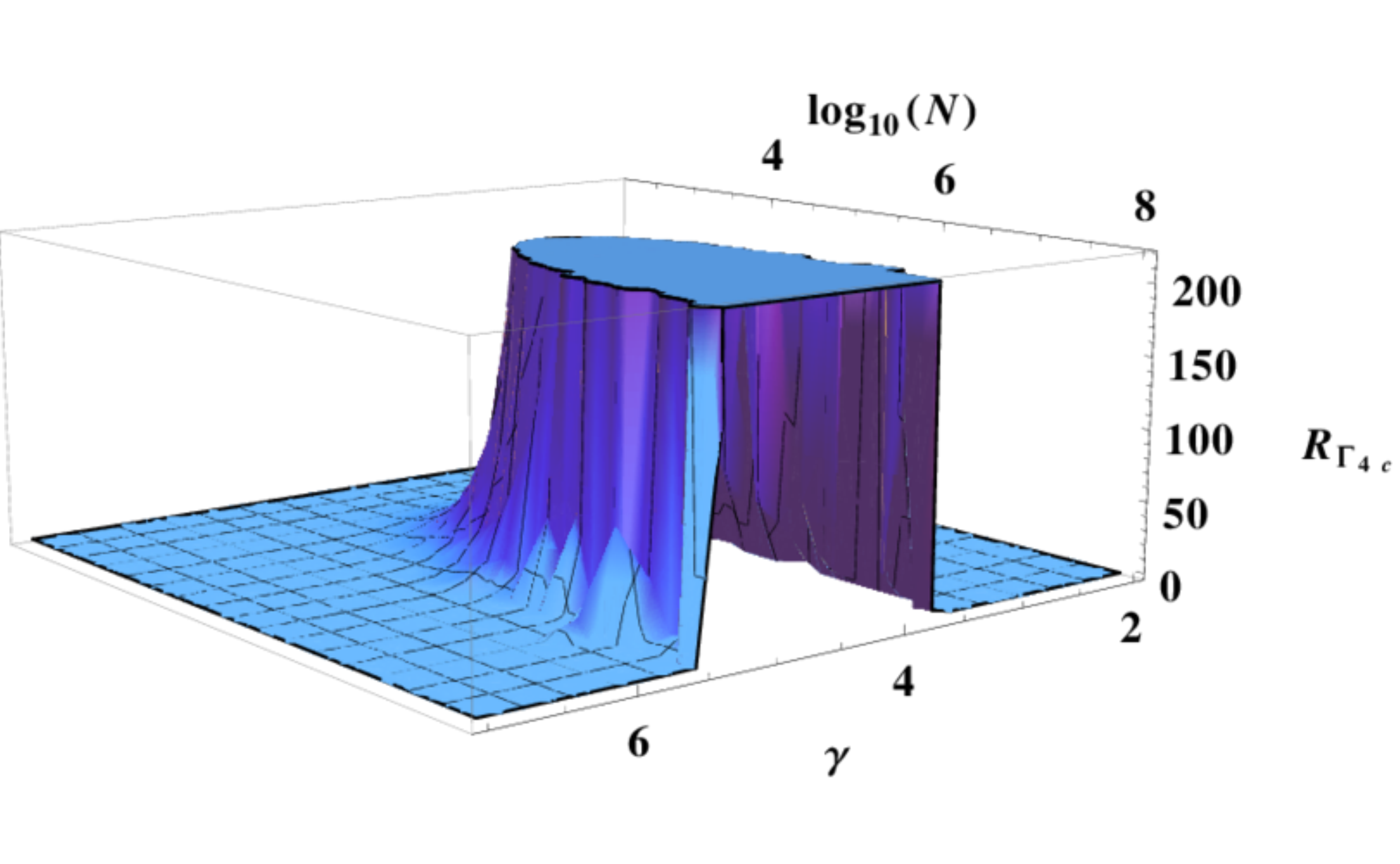}}
%{\includegraphics[height=3.5in]{figures/fig4_PRE.pdf}}
\caption{Behavior of $R^{(A)}_{\Gamma_{4c}}$, as a function of $\gamma$ and $N$,
obtained from numerical integration of Eq. (\ref{RcliqueA}) with $k=4$ by using $10^7$ points per integral. 
\label{fig4}
}
\end{figure}
\begin{figure}[tbh]
{\includegraphics[height=3.5in]{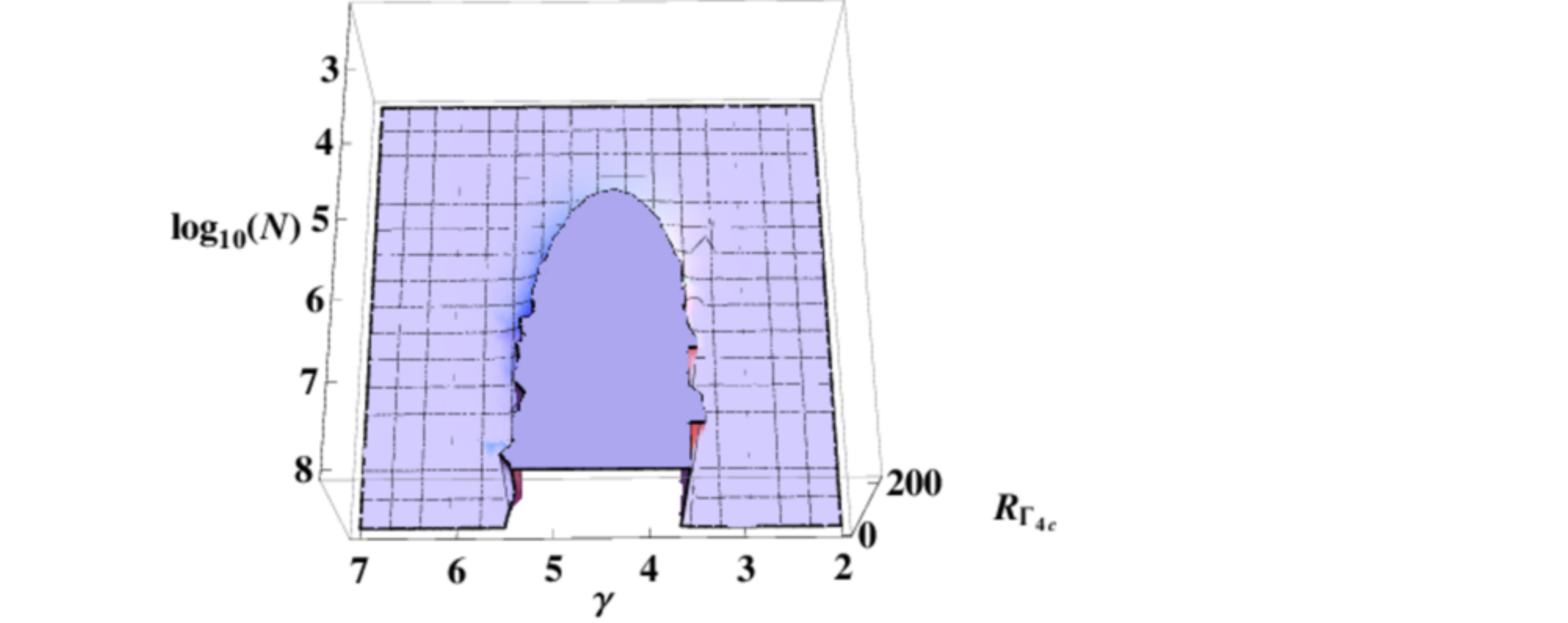}}
%{\includegraphics[height=3.5in]{figures/fig5_PRE.pdf}}
\caption{Top view of Fig. \ref{fig4}.
\label{fig5}
}
\end{figure}
\begin{figure}[tbh]
{\includegraphics[height=3.5in]{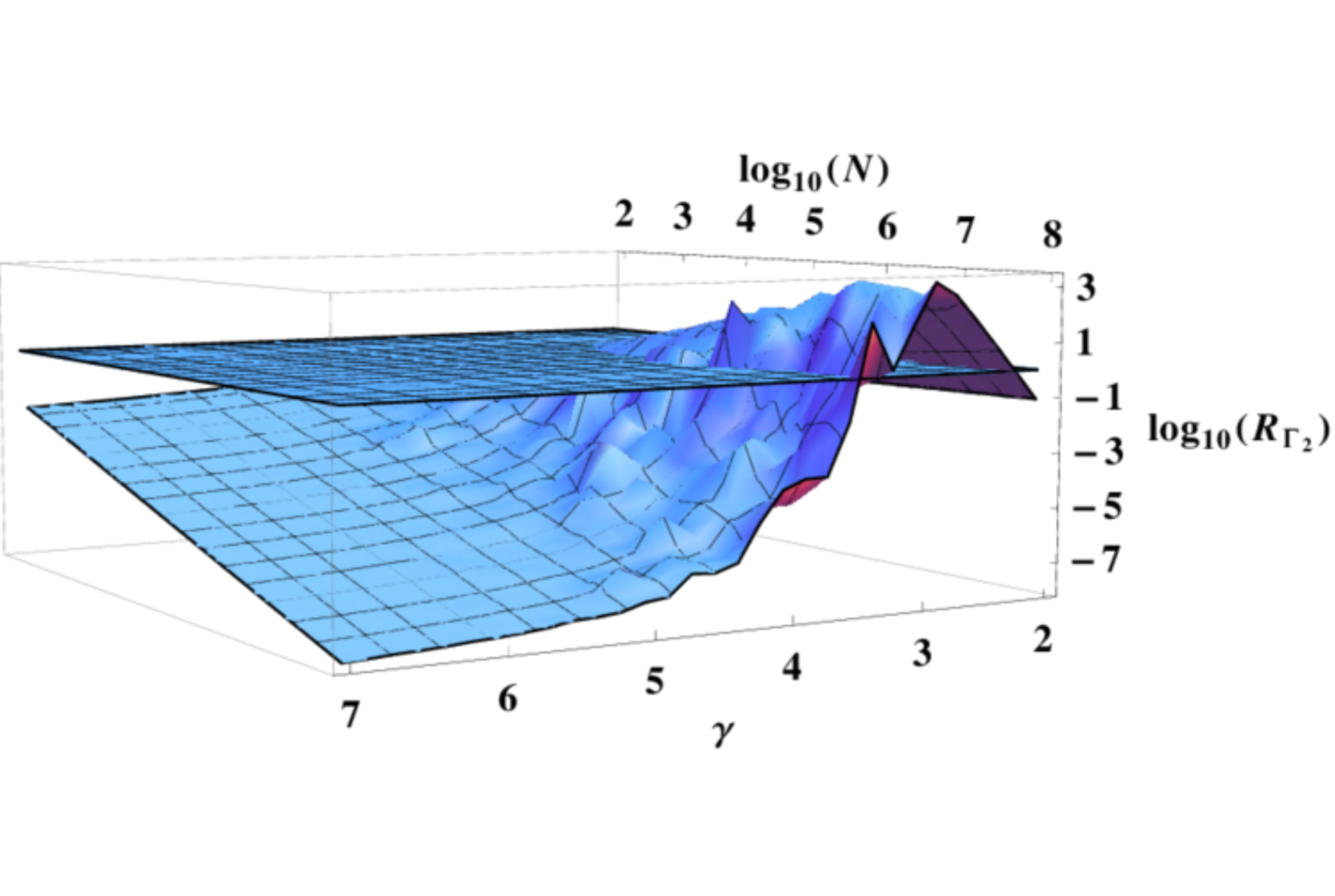}}
%{\includegraphics[height=3.5in]{figures/fig6_PRE.pdf}}
\caption{Behavior of $R^{(A)}_{\Gamma_{2}}$,  as a function of $\gamma$ and $N$,
obtained from numerical integration of Eq. (\ref{RTRA}) by using $10^7$ points per integral.
In this case, a higher number of points per integral would be desirable in order to reduce the irregularities
present in this figure. 
%The necessity of a relatively higher statistics is due to the fact that $R^{(A)}_{\Gamma_{2}}$,
%compared to $R^{(A)}_{\Gamma_{3}}$ or $R^{(A)}_{\Gamma_{4c}}$, diverges in a smaller window of $\gamma$. 
Notice however that, in this case, for $R^{(A)}_{\Gamma_{2}}$ we have a logarithmic scale.
\label{fig6}
}
\end{figure}
\begin{figure}[tbh]
{\includegraphics[height=3.5in]{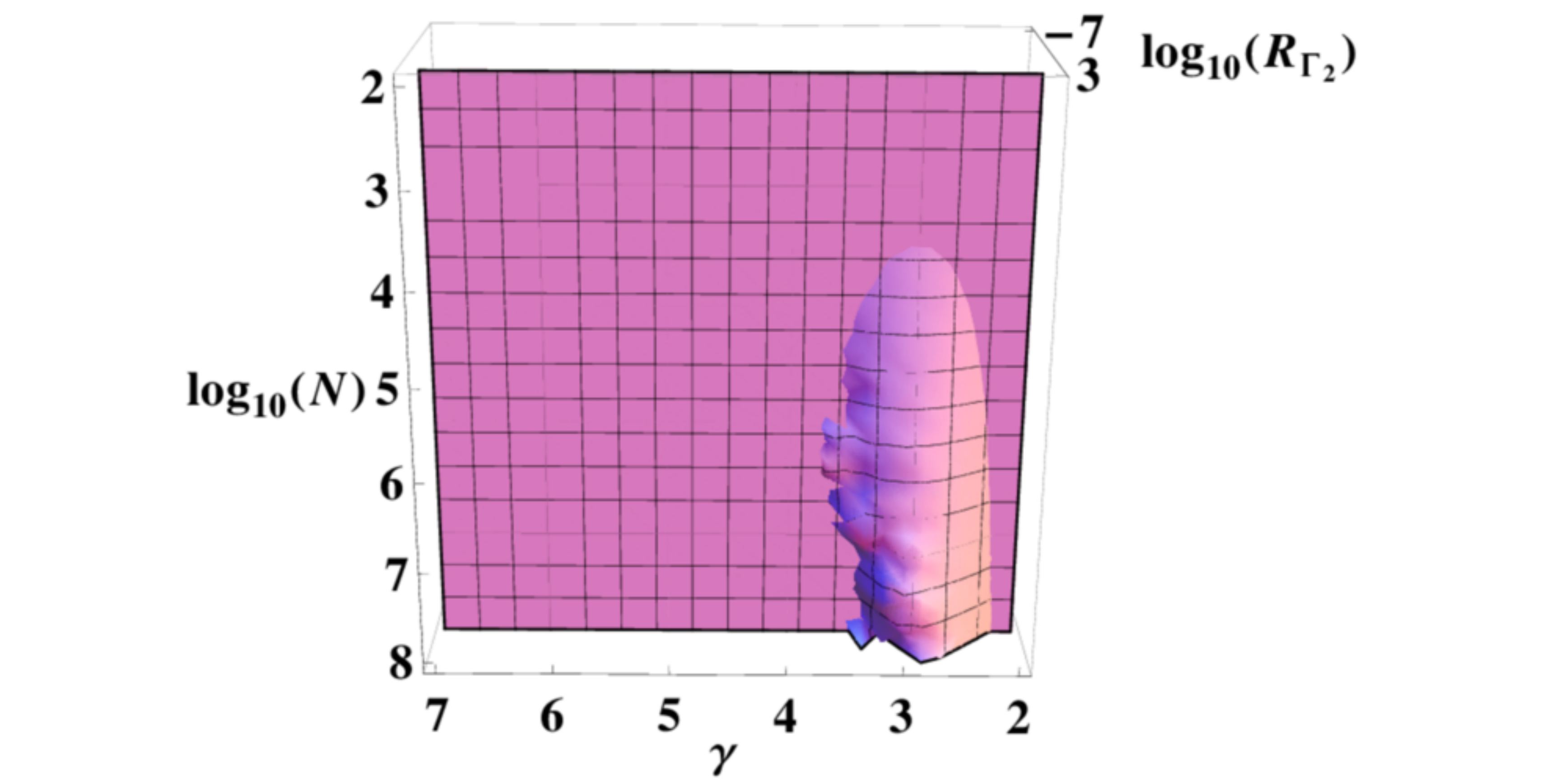}}
%{\includegraphics[height=3.5in]{figures/fig7_PRE.pdf}}
\caption{Top view of Fig. \ref{fig6}.
\label{fig7}
}
\end{figure}
\begin{figure}[tbh]
{\includegraphics[height=3.5in]{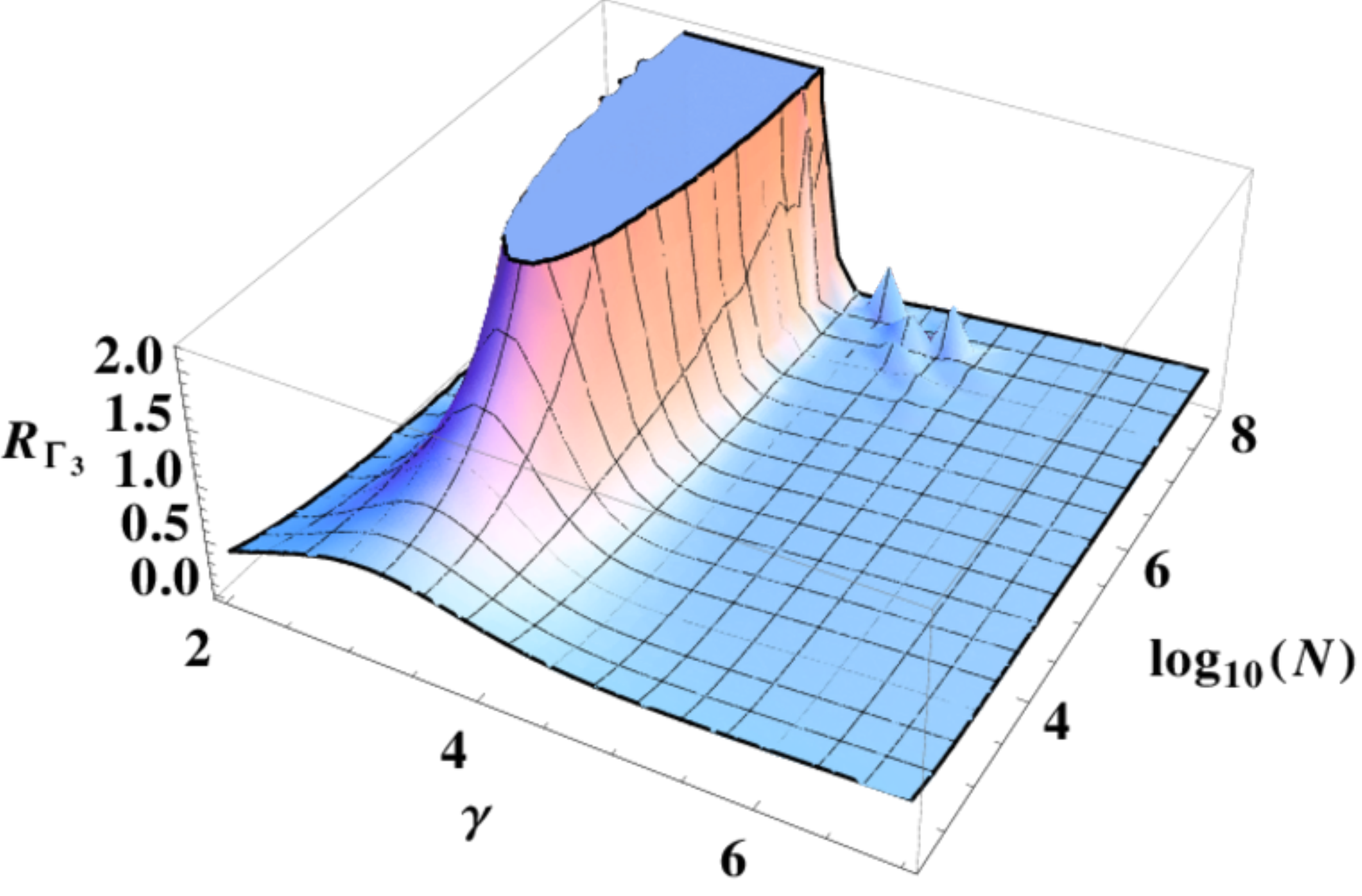}}
%{\includegraphics[height=3.5in]{figures/fig2_PRE.pdf}}
\caption{Behavior of $R^{(S)}_{\Gamma_3}$, as a function of $\gamma$ and $N$,
obtained from numerical integration of Eq. (\ref{RTS}) by using $10^7$ points per integral. 
Also here, a higher number of points per integral would be desirable in order to reduce the small irregularities
that, on this scale, turn out to be visible. 
\label{fig14}
}
\end{figure}
\begin{figure}[tbh]
{\includegraphics[height=3.5in]{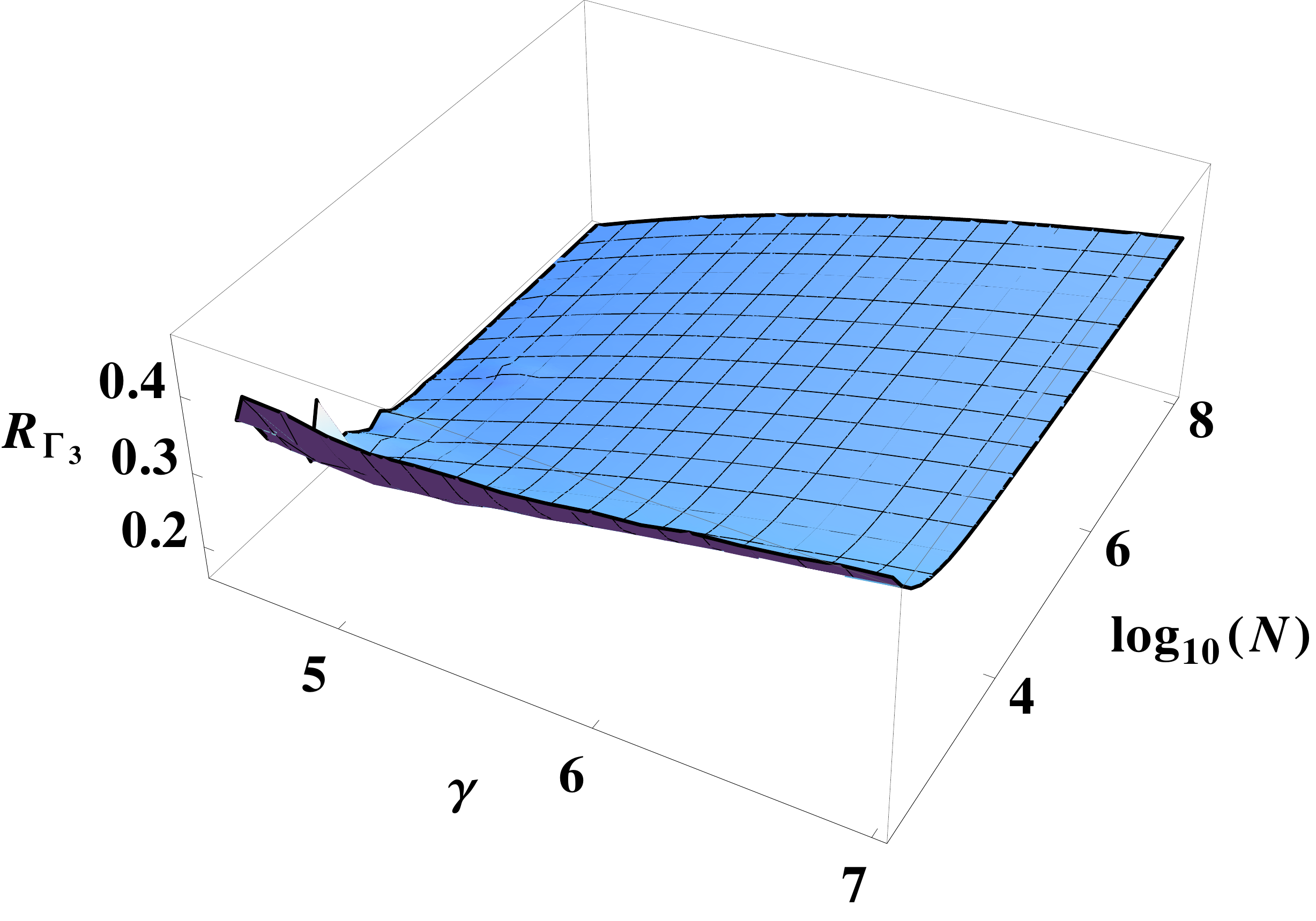}}
%{\includegraphics[height=3.5in]{figures/fig2_PRE.pdf}}
\caption{Particular of fig. \ref{fig14}. On comparing with Figs. \ref{fig2} and \ref{fig14}, 
it makes evident that, whereas  $R^{(A)}_{\Gamma_3}$ and $R^{(S)}_{\Gamma_3}$ turn out to be similar
for $\gamma \in [\gamma_1,\gamma_2]$, they are quite different for larger value of $\gamma$. In fact,
$\lim_{\gamma\to\infty}R^{(A)}_{\Gamma_3}= 0$, while $\lim_{\gamma\to\infty}R^{(S)}_{\Gamma_3}\neq 0$. See Eqs. (\ref{RSAM}-\ref{RqM}).
\label{fig15}
}
\end{figure}
\begin{figure}[tbh]
{\includegraphics[height=3.5in]{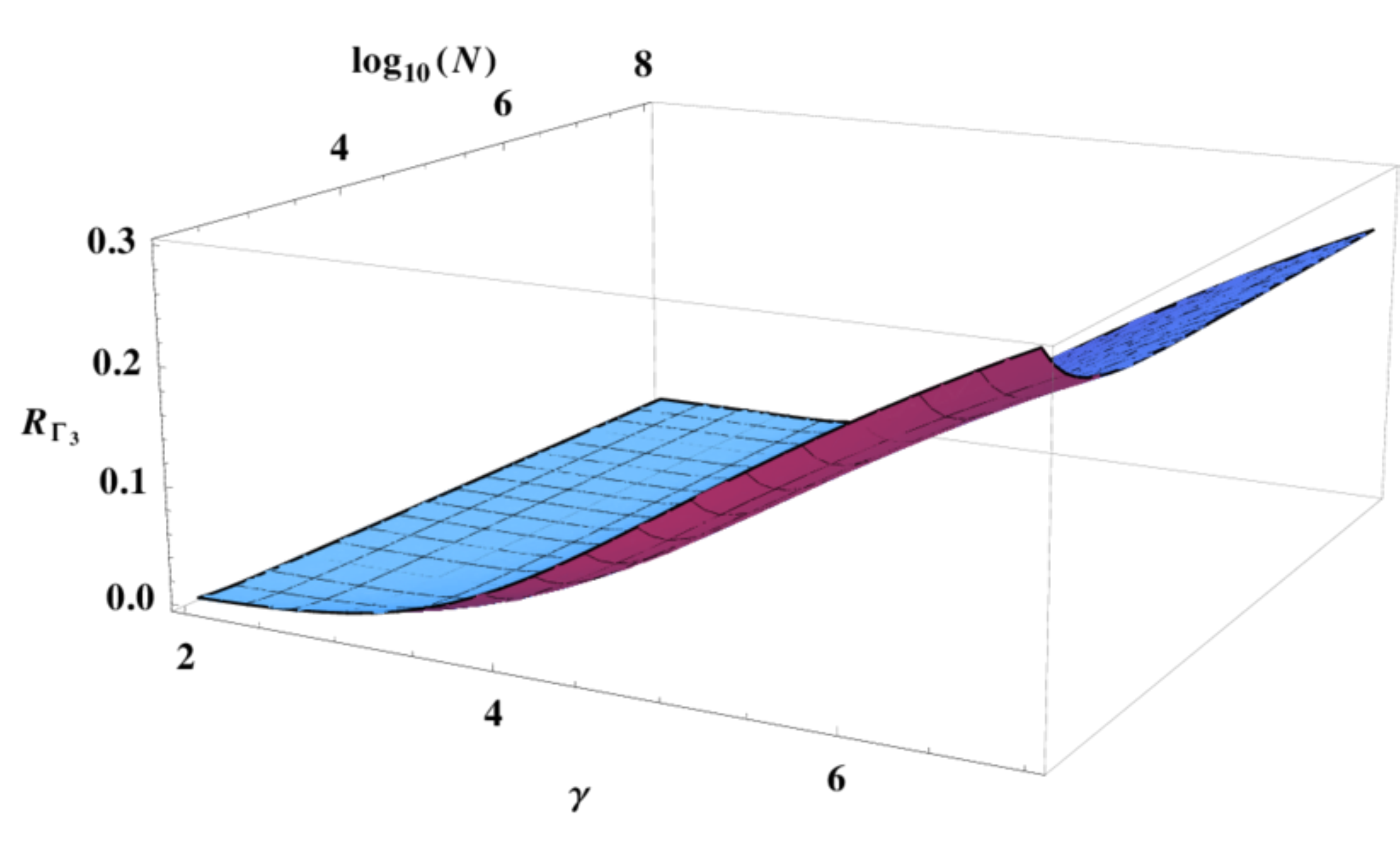}}
%{\includegraphics[height=2.5in]{figures/fig8_PRE.pdf}}
\caption{Behavior of $R^{(B)}_{\Gamma_{3}}$, as a function of $\gamma$ and $N$,
obtained from numerical integration of Eq. (\ref{RTB}) by using $10^7$ points per integral. 
On comparing with the previous figures we clearly see the behavior expressed by Eqs. (\ref{RBM}-\ref{RSABM1}).
\label{fig8}
}
\end{figure}

\end{widetext}
Here we list the main properties of the standard ratios $R^{(\mathrm{X})}_\mathsmaller{\Gamma}$ in the five above ensembles.
The proofs of these Eqs., as well as the detailed evaluation of $R^{(\mathrm{X})}_\mathsmaller{\Gamma}$ for specific
motifs, will be given in the next Sections.
Let $\Gamma$ be any motif 
having minimal and maximal degree $k_{\mathrm{min}}$ and $k_{\mathrm{max}}$, respectively.
There exist $\gamma_1$ and $\gamma_2$ with
$\gamma_1\approx \max\{2,k_{\mathrm{min}}\}$ and $\gamma_2\approx 2 k_{\mathrm{max}}$, such that
\begin{eqnarray}
\label{RSAM}
\lim_{N \to \infty} R^{(\mathrm{S})}_\mathsmaller{\Gamma}= \lim_{N \to \infty} R^{(\mathrm{A})}_\mathsmaller{\Gamma}=\infty, \quad \mathrm{if~} \gamma\in (\gamma_1,\gamma_2), 
\end{eqnarray}
\begin{eqnarray}
\label{RSAM1}
\lim_{N \to \infty} R^{(\mathrm{S})}_\mathsmaller{\Gamma}= \lim_{N \to \infty} R^{(\mathrm{A})}_\mathsmaller{\Gamma}=0, \quad \mathrm{if~} \gamma\notin (\gamma_1,\gamma_2), 
\end{eqnarray}
\begin{eqnarray}
\label{RBM}
\lim_{N \to \infty} R^{(\mathrm{B})}_\mathsmaller{\Gamma}=\mathrm{O}(1), \quad \forall \gamma<\infty
\end{eqnarray}
\begin{eqnarray}
\label{RSABM}
\lim_{\gamma \to \infty} R^{(\mathrm{S})}_\mathsmaller{\Gamma}\neq 0, \quad \lim_{\gamma \to \infty} R^{(\mathrm{A})}_\mathsmaller{\Gamma}= 0, 
\quad \lim_{\gamma \to \infty}  R^{(\mathrm{B})}_\mathsmaller{\Gamma} \neq 0,
\end{eqnarray}
%\begin{eqnarray}
%\label{RAM}
%\lim_{\gamma \to \infty} R^{(\mathrm{A})}_\mathsmaller{\Gamma}= 0,
%\end{eqnarray}
\begin{eqnarray}
\label{RSABM1}
R^{(\mathrm{S})}_\mathsmaller{\Gamma} > R^{(\mathrm{A})}_\mathsmaller{\Gamma}, \quad R^{(\mathrm{S})}_\mathsmaller{\Gamma} > R^{(\mathrm{B})}_\mathsmaller{\Gamma}, 
\end{eqnarray}
\begin{eqnarray}
\label{RhM}
R^{(\mathrm{Q})}_{\mathsmaller{\Gamma}|\bm{h}}=\mathrm{O}\left(\frac{1}{N}\right),
\end{eqnarray}
\begin{eqnarray}
\label{RqM}
R^{(\mathrm{Q})}_{\mathsmaller{\Gamma}|\bm{q}}=\mathrm{O}\left(R^{(\mathrm{S})}_\mathsmaller{\Gamma}\right)=\mathrm{O}\left(R^{(\mathrm{A})}_\mathsmaller{\Gamma}\right), \quad \forall \gamma<\infty. 
\end{eqnarray}

Eqs. (\ref{RSAM}) and (\ref{RSAM1}) encode the phase transition scenario for the ensembles (S) and (A): if the motif $\Gamma$ is such that
$\gamma\notin(\gamma_1,\gamma_2)$, the density $n_{_\Gamma}$ is self-averaging, otherwise, if $\gamma\in(\gamma_1,\gamma_2)$,
$n_{_\Gamma}$ is non-self-averaging or, more precisely, strongly non-self-averaging (see Figs. \ref{fig2}-\ref{fig14}).
Eqs. (\ref{RBM}) tells us that the hidden-variables responsible for the diverging behavior of $R^{(\mathrm{S})}_\mathsmaller{\Gamma}$ are only 
the node-hidden-variables $h$'s, whereas the link-hidden-variables $q$'s do not produce important fluctuations (see Fig. \ref{fig8}). In fact, Eq. (\ref{RhM})
tells us that the $q$'s lead always to the standard law of large numbers. This is perfectly compatible with the fact that,
if the $h$' s are fixed parameters, the degree distribution of a node with fixed node-hidden-variable is Poissonian \cite{BogunaHidden},
and since Poissonian fluctuations are small, the law of large numbers applies.
As a consequence of Eqs. (\ref{RSAM})-(\ref{RBM}) we obtain also Eq. (\ref{RqM}).
Eqs. (\ref{RSABM}) establish that only by first sampling the $q$'s and then averaging the result over the $h$' s,
we can recover the total absence of correlations, intuitively expected when $\gamma\to\infty$, which is the classical random graph limit.
Finally, Eqs. (\ref{RSABM1}) provide useful inequalities.

As we had anticipated in the Introduction, another consequence of the non-self-averaging scenario concerns the practical problems occurring when the sequence 
of the expected degrees $\{h_i\}$ is supposed to be fixed.
We have to take into account that, on the base of a given real-network, the expected degree sequence $\{h_i\}$ must be measured and
it is affected by unavoidable finite errors $\{\epsilon_i\}$.
Now, Eqs. (\ref{RSAM}) and (\ref{RhM}) imply that, if we are in a non-self-averaging region, such errors, even if small, in general might seriously affect the network resulting from the hidden-variable model. 
In fact, if we call $h_i$ the values of the real network, and $h_i'=h_i+\epsilon_i$ the values of the measured expected degrees, by using integration by parts together with $\rho(h)\to 0$ for $h\to \infty$, and $p(h,h')\to 0$ for $h\to \infty$ or $h'\to\infty$, for any motif $\Gamma$ we have
\begin{eqnarray}
\label{DD}
&& \frac{\mediaS{\left(\mediaT{n_\mathsmaller{\Gamma}}_{|_{\bm{h'}}}-\mediaT{n_\mathsmaller{\Gamma}}_{|_{\bm{h}}}\right)^2}}
{\mediaS{\mediaT{n_\mathsmaller{\Gamma}}_{|_{\bm{h}}}}^2} \simeq
\frac{-\sum_i \epsilon_i \mediaS{\frac{\partial}{\partial h_i}\mediaT{n_\mathsmaller{\Gamma}}^2_{|_{\bm{h}}}}}
{\mediaS{\mediaT{n_\mathsmaller{\Gamma}}_{|_{\bm{h}}}}^2} \nonumber \\
&\simeq& \gamma \frac{a}{a_1} \left(1+R^{(\mathrm{A})}_{\Gamma}\right)\sum_i \epsilon_i, 
\end{eqnarray}
where: $a$ is given by Eq. (\ref{a}), $a_1$ is given by Eq. (\ref{a}) evaluated for $\gamma\to \gamma-1$, 
and the above approximations become exact in the limit 
$N\to\infty$ and $\{\epsilon_i\to 0\}$.
In general $R^{(\mathrm{A})}_{\Gamma}$ diverges for $N\to\infty$ when $\gamma\in (\gamma_1(\Gamma),\gamma_2(\Gamma))$.
As a consequence, from Eq. (\ref{DD}), we see that even small errors in the measure of the expected degree may generate large differences between the real- and the simulated-network. 

The next Section is devoted to the proofs of Eqs. (\ref{RSAM})-(\ref{RqM}), and to a detailed evaluation of $R^{(\mathrm{X})}_\mathsmaller{\Gamma}$ for specific
motifs. 

\section{Analysis of $R^{(\mathrm{X})}$}
Given a motif $\Gamma$, the density of  $\Gamma$ in a graph realization is
\begin{eqnarray}
\label{nM}
%n_\Gamma=\frac{c}{N}\sum_i \chi_{i}^{(\Gamma)},
n_\mathsmaller{\Gamma}=\frac{c}{N}\sum_i k_{\mathsmaller{\Gamma}}(i),
\end{eqnarray}
where $k_{\mathsmaller{\Gamma}}(i)$ counts the number of motifs $\Gamma$ passing through the node $i$. 
The coefficient $c$ depends on the definition
of the motif considered and serves to avoid over-counting when the motif is symmetric. For example,
if the motif $\Gamma$ is the triangle, we set $c=1/3$. If the motif is not symmetric, we can establish to count
only those motifs that pass through a specif node of $\Gamma$. For example, if $\Gamma$ is a triple (two consecutive links), 
we can set $c=1$, but a motif contributes only when the center of the triple coincides with $i$.
However, since we are interested only in the relative fluctuations $R^{(\mathrm{X})}_\mathsmaller{\Gamma}$, we do not need to specify it since $c$, as well as any other constant, 
does not play any role for $R^{(\mathrm{X})}_\mathsmaller{\Gamma}$. Let us consider now the numerators of Eqs. (\ref{RS}-\ref{Rq}). 
Note that the hidden variable scheme does not distinguish nodes, therefore we 
can make use of the fact that nodes, once averaged, are all statistically equivalent. 
By using this property, from Eq. (\ref{nM}) we get the following susceptibilities (note that we reserve the symbols $\mediaS{\cdot}$ and $\mediaG{\cdot}$
to the averages over the $h$'s and the $q$'s respectively and, therefore, to avoid ambiguities, for arithmetical parenthesis we will use only the symbol $(\cdot)$)
%By plugging Eq. (\ref{nM}) into the numerator of Eq. (\ref{R}), and 
%by using the statical equivalence of nodes, we get the following susceptibility
%\begin{eqnarray}
%\label{RN0}
%&& \mediaT{{n_\mathsmaller{\Gamma}}^2}-\mediaT{n_\mathsmaller{\Gamma}}^2=\nonumber \\
%&& \frac{c^2\sum_{i,j}\left\{\mediaT{k_{\mathsmaller{\Gamma}}(i)k_{\mathsmaller{\Gamma}}(j)}-\mediaT{k_{\mathsmaller{\Gamma}}(i)}\mediaT{k_{\mathsmaller{\Gamma}}(j)}\right\}}{N^2},
%\end{eqnarray}
\begin{widetext}
\begin{eqnarray}
\label{RNS}
&& \mediaT{{n_\mathsmaller{\Gamma}}^2}-\mediaT{n_\mathsmaller{\Gamma}}^2=
\frac{c^2}{N}\left(\mediaT{k^2_{\mathsmaller{\Gamma}}(i)}-\mediaT{k_{\mathsmaller{\Gamma}}(i)}^2\right)
+ c^2\left(\mediaT{k_{\mathsmaller{\Gamma}}(i)k_{\mathsmaller{\Gamma}}(j)}_{i\neq j}-\mediaT{k_{\mathsmaller{\Gamma}}(i)}^2\right),
\end{eqnarray}
\begin{eqnarray}
\label{RNA}
\mediaS{\mediaT{\mathop{n_\mathsmaller{\Gamma}}}^2_{|\bm{h}}}-\mediaS{\mediaT{\mathop{n_\mathsmaller{\Gamma}}}_{|\bm{h}}}^2=
\frac{c^2}{N}\left(\mediaS{\mediaT{k_{\mathsmaller{\Gamma}}(i)}_{|\bm{h}}^2}-\mediaS{\mediaT{k_{\mathsmaller{\Gamma}}(i)}_{|\bm{h}}}^2\right)
+ c^2\left(\mediaS{\mediaT{k_{\mathsmaller{\Gamma}}(i)}\mediaT{k_{\mathsmaller{\Gamma}}(j)}_{i\neq j~|\bm{h}}}-\mediaS{\mediaT{k_{\mathsmaller{\Gamma}}(i)}_{|\bm{h}}}^2\right),
\end{eqnarray}
\begin{eqnarray}
\label{RNB}
\mediaG{\mediaT{\mathop{n_\mathsmaller{\Gamma}}}^2_{|\bm{q}}}-\mediaG{\mediaT{\mathop{n_\mathsmaller{\Gamma}}}_{|\bm{q}}}^2=
\frac{c^2}{N}\left(\mediaG{\mediaT{k_{\mathsmaller{\Gamma}}(i)}_{|\bm{q}}^2}-\mediaG{\mediaT{k_{\mathsmaller{\Gamma}}(i)}_{|\bm{q}}}^2\right)
+ c^2\left(\mediaG{\mediaT{k_{\mathsmaller{\Gamma}}(i)}\mediaT{k_{\mathsmaller{\Gamma}}(j)}_{i\neq j~|\bm{q}}}-\mediaG{\mediaT{k_{\mathsmaller{\Gamma}}(i)}_{|\bm{q}}}^2\right),
\end{eqnarray}
\begin{eqnarray}
\label{RNQA}
\mediaS{\mediaT{\mathop{n_\mathsmaller{\Gamma}}^2}_{|\bm{h}}}-\mediaS{\mediaT{\mathop{n_\mathsmaller{\Gamma}}}^2_{|\bm{h}}}=
\frac{c^2}{N}\left(\mediaS{\mediaT{k_{\mathsmaller{\Gamma}}(i)^2}_{|\bm{h}}}-\mediaS{\mediaT{k_{\mathsmaller{\Gamma}}(i)}_{|\bm{h}}^2}\right)
+ c^2\left(\mediaS{\mediaT{k_{\mathsmaller{\Gamma}}(i)k_{\mathsmaller{\Gamma}}(j)}_{i\neq j~|\bm{h}}}-\mediaS{\mediaT{k_{\mathsmaller{\Gamma}}(i)}_{|\bm{h}}^2}\right),
\end{eqnarray}
\begin{eqnarray}
\label{RNQB}
\mediaG{\mediaT{\mathop{n_\mathsmaller{\Gamma}}^2}_{|\bm{q}}}-\mediaG{\mediaT{\mathop{n_\mathsmaller{\Gamma}}}_{|\bm{q}}^2}=
\frac{c^2}{N}\left(\mediaG{\mediaT{k_{\mathsmaller{\Gamma}}(i)^2}_{|\bm{q}}}-\mediaG{\mediaT{k_{\mathsmaller{\Gamma}}(i)}_{|\bm{q}}^2}\right)
+ c^2\left(\mediaG{\mediaT{k_{\mathsmaller{\Gamma}}(i)k_{\mathsmaller{\Gamma}}(j)}_{i\neq j~|\bm{q}}}-\mediaG{\mediaT{k_{\mathsmaller{\Gamma}}(i)}_{|\bm{q}}^2}\right),
\end{eqnarray}
\end{widetext}
where $i$ and $j$ represent two arbitrary distinct indices. 
In the rhs of Eqs. (\ref{RNS}-\ref{RNQB}) we have approximated $(N-1)/N$ to 1. As for the susceptibility of an homogeneous system,
for each of the Eqs. (\ref{RNS}-\ref{RNQB}), in the rhs,
we have a self-term proportional to the motif variance, 
%(the variance for the case in which $\Gamma$ is a link is well known)
rescaled by the factor $1/N$,
and a mixed-term that accounts for correlations between two motifs centered at two different nodes.
%The prefactor $(N-1)/N$ that appeares in the mixed-term can always be approximated to 1, but we anticipate that the self-term, despite appears to be order $1/N$, in general cannot be
Note that, the self-term, despite appears to be order $1/N$, in general cannot be
neglected. In fact, due to exact cancellations in the mixed term, the mixed- and self-terms give contributions of the same order of magnitude.  
We anticipate that the mixed term is always zero in Eq. (\ref{RNQA}).

In the following, by using Eqs. (\ref{RNA}-\ref{RNQB}), we analyze $R^{(\mathrm{X})}$ for each X-ensemble for a few crucial motifs and then we extrapolate the general behavior.

\subsection{Diagrammatic calculus}
One of the main advantage of hidden-variable models lies in the fact that we can reduce the average of almost any observable to suitable integrals.
As will see later on, this is particularly true for the evaluation of the motif densities and their fluctuations.
In fact, a correspondence between formulas and diagrams can be established to avoid unnecessary simulations and to improve
our understanding about the main contributions to the fluctuations, especially those that can generate non self-averaging. 
We then introduce the compact notation $\mediaS{\Gamma}$, where $\Gamma$ can be any motif.
For example, by referring to Fig. \ref{fig1}, we have 
%$\mediaS{\Gamma_1}=\mediaS{p(h_1,h_2)}$, $\mediaS{\Gamma_2}=\mediaS{p(h_1,h_2)p(h_2,h_3)}$,
%$\mediaS{\Gamma_3}=\mediaS{p(h_1,h_2)p(h_2,h_3)p(h_3,h_1)}$,
\begin{eqnarray}
\label{G3l}
&&\mediaS{\Gamma_{3l}}=\mediaS{p(h_1,h_2)p(h_2,h_3)p(h_3,h_4)},
\end{eqnarray}
\begin{eqnarray}
\label{G3}
&&\mediaS{\Gamma_3}=\mediaS{p(h_1,h_2)p(h_2,h_3)p(h_3,h_1)},
\end{eqnarray}
\begin{eqnarray}
\label{G3c}
&&\mediaS{\Gamma_{3s}}=\mediaS{p(h_4,h_1)p(h_4,h_2)p(h_4,h_3)},
\end{eqnarray}
and so on. The role played by these $\mediaS{\cdot}$-averages, is very similar
to the role played by the Green functions in statistical field theory. 
Moreover, if we are interested in the evaluation 
of connected correlation functions, we need to work only with Green functions of connected motifs, as the contributions
of disconnected motifs always cancel. 
The ``degree'' of correlation of a given hidden-variable model is encoded in these integrals 
to be evaluated numerically. Only in the limit $\gamma\to\infty$ the correlations become negligible
(or more precisely logarithmically negligible, with corrections that scales as $1/\log N$). In fact, as a general rule we have 
\begin{eqnarray}
\label{gammal}
&& \lim_{\gamma\gg 1}\mediaS{p(h_1,h_2)p(h_2,h_3)\ldots p(h_m,h_{m+1})}=\nonumber \\
&& \mediaS{p(h_1,h_2)}\mediaS{p(h_2,h_3)}\ldots \mediaS{p(h_m,h_{m+1})}.
\end{eqnarray}
Eq. (\ref{gammal}) tells us that, in the limit of large $\gamma$, \textit{i.e.}, the limit
where the network becomes indistinguishable from the classical random graph, 
a connected motif becomes indistinguishable from the set of its disconnected components and, as a consequence, 
the connected correlation functions tend to zero. 
As will see better later, Eq. (\ref{gammal}) provides us also a useful tool to have a simple estimate 
of the averages of the density of motifs (but not of its variance). 

\subsection{$R^{(\mathrm{S})}$ (Sampling the hidden-variables symmetrically)}
Here we consider the most common case in which we sample the two set of hidden-variables symmetrically and we have to
evaluate the fluctuations thorough Eq. (\ref{RNS}). 
This symmetric sampling has already been analyzed in \cite{SAEPL}. Here we review this case providing
further details which were omitted in \cite{SAEPL}.

\begin{figure}[tbh]
{\includegraphics[height=1.6in]{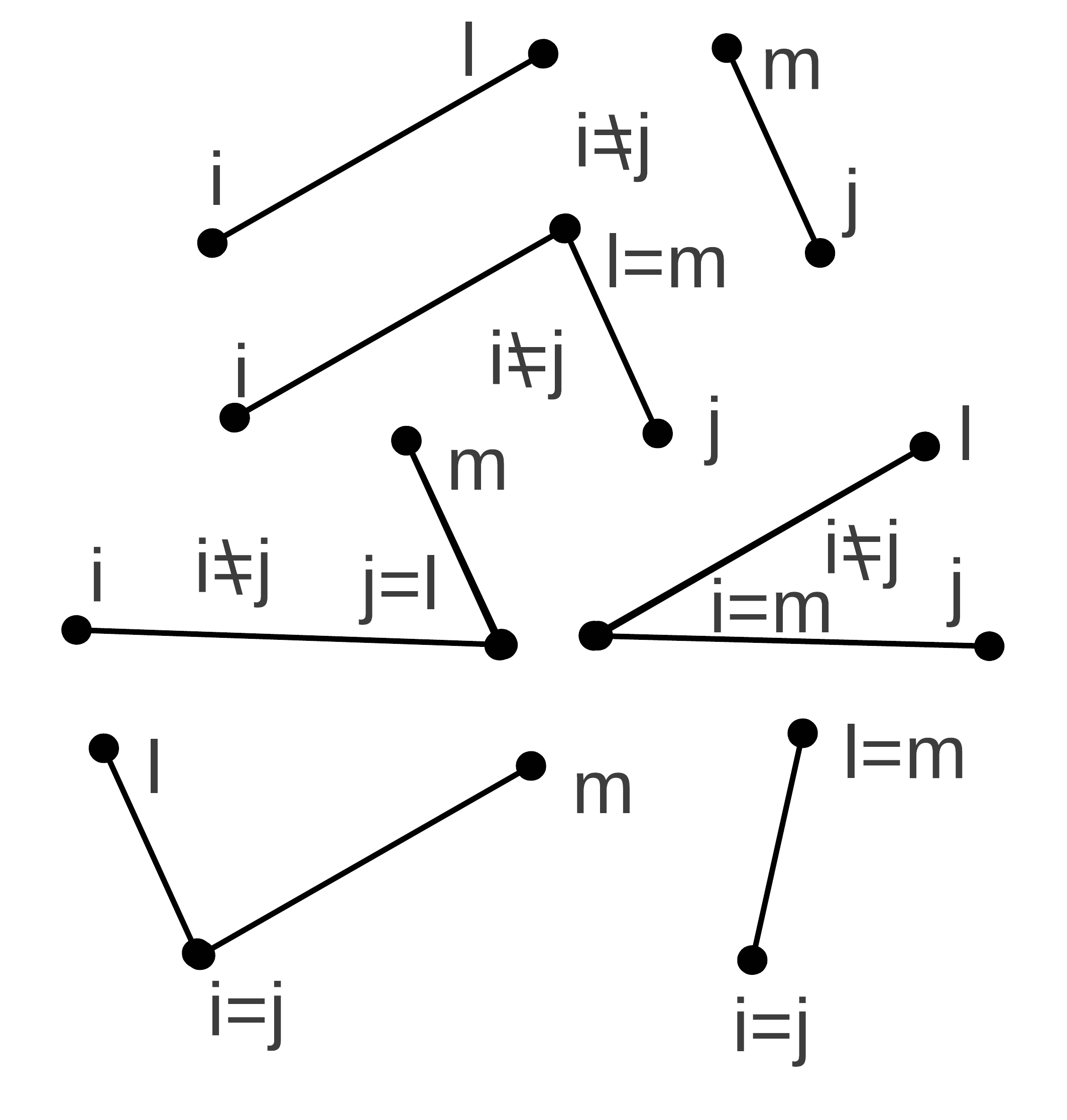}}
%{\includegraphics[height=1.6in]{figures/fig9_PRE.pdf}}
\caption{Contributions to Eqs. (\ref{link3}) (lower connected motifs) and (\ref{link4}) (upper disconnected motifs and the
3 connected motifs located in the central part of the figure). 
Nodes $i$ and $j$ are to be kept fixed, while all the others can vary,
provided the topology is kept fixed.
Contributions coming from disconnected motifs always cancel in $R_\mathsmaller{\Gamma}$.
\label{fig9}
}
\end{figure}

\textit{Link} $(\Gamma_{1})$.
If $\Gamma$ is the link, $k_{\mathsmaller{\Gamma}}(i)$ coincides
with the standard definition of degree of the node $i$. 
For a given graph realization, corresponding to a given realization of the $h$'s, in terms of adjacency matrix, we have
\begin{eqnarray}
\label{link}
k_{\mathsmaller{\Gamma_{1}}}(i)=\sum_{l\neq i}a_{i,l}.
\end{eqnarray}
By using Eqs. (\ref{CM})-(\ref{aveh}), and the statistical equivalence of nodes, we have
\begin{eqnarray}
\label{link1}
\mediaT{k_{\mathsmaller{\Gamma_{1}}}(i)}=\sum_{l\neq i}\mediaT{a_{i,l}}=(N-1)\mediaS{p(h_1,h_2)}.
\end{eqnarray}
Let us now consider the product $k_{\mathsmaller{\Gamma_{1}}}(i)k_{\mathsmaller{\Gamma_{1}}}(j)$.
%\begin{eqnarray}
%\label{link2}
%k_{\mathsmaller{\Gamma_{1}}}(i)k_{\mathsmaller{\Gamma_{1}}}(j)=\sum_{l\neq i,m\neq j}a_{i,l}a_{j,m}.
%%\mediaT{k_{\mathsmaller{\Gamma}}(i)k_{\mathsmaller{\Gamma}}(j)}=\sum_{l\neq i,m\neq j}p(h_i,h_l)p(h_j,h_m).
%\end{eqnarray}
Notice that $a^2_{i,j}=a_{i,j}$.
We have to distinguish the case $i=j$ and $i\neq j$, see Fig.~\ref{fig9}. 
For $i=j$ we have
\begin{eqnarray}
\label{link3}
&&\mediaT{k^2_{\mathsmaller{\Gamma_{1}}}(i)}%=\sum_{l,m\neq i}\mediaS{p(h_i,h_l)p(h_i,h_m)}+\sum_{l\neq i}\mediaS{p(h_i,h_l)} \nonumber\\ && 
=(N-1)(N-2)\mediaS{p(h_1,h_2)p(h_2,h_3)}
\nonumber\\ && + (N-1)\mediaS{p(h_1,h_2)},
\end{eqnarray}
while for $i\neq j$ we have
\begin{eqnarray}
\label{link4}
&& \mediaT{k_{\mathsmaller{\Gamma_{1}}}(i)k_{\mathsmaller{\Gamma_{1}}}(j)}%=\sum_{l\neq i,m\neq j:l\neq m}\mediaS{p(h_i,h_l)p(h_j,h_m)} 
%\nonumber\\ && + 3 \sum_{l\neq i,j}\mediaS{p(h_i,h_l)p(h_l,h_j)}=
=(N-2)(N-3)\mediaS{p(h_1,h_2)}^2
\nonumber\\ && + 3(N-2)\mediaS{p(h_1,h_2)p(h_2,h_3)},
\end{eqnarray}
where the factor 3 in the last term comes from the fact that two links emanating from nodes $i$ and $j$ can share a same node in 3 
topologically equivalent ways. We finally plug Eqs. (\ref{link1})-(\ref{link4}) into Eq. (\ref{RS}) via Eq. (\ref{RNS}) and, 
by keeping in this latter only terms in $N^2$, which cancel exactly, and terms in $N$, we obtain
\begin{eqnarray}
\label{RLS}
R^{(\mathrm{S})}_\mathsmaller{\Gamma_{1}}=\frac{4}{N}\frac{\mediaS{p(h_1,h_2)p(h_2,h_3)}}{\mediaS{p(h_1,h_2)}^2}-\frac{4}{N}+\frac{1}{N^2\mediaS{p(h_1,h_2)}}.
\end{eqnarray}
Due to Eq. (\ref{avea}) and its generalizations, for $\gamma>3$, each term present in the rhs of Eq. (\ref{RLS}) is of order $1/N$, so that
we have also $R^{(\mathrm{S})}_\mathsmaller{\Gamma_{1}}=\mathrm{O}(1/N)$ and the network is self-averaging with respect to the link density, 
while for $\gamma<3$ we have still self-averaging but $R^{(\mathrm{S})}_\mathsmaller{\Gamma_{1}}$ decays slower as $N^{2-\gamma}$. 
It is interesting however to observe the general behavior of $R_\mathsmaller{\Gamma_{1}}$ with respect to $\gamma$ for finite $N$.   
Due to Eq. (\ref{gammal}) the first two terms in the rhs of Eq. (\ref{RLS}) tend to cancel for large $\gamma$. However, the last term does not cancel
for large $\gamma$. We will see that in the A ensemble this term instead cancels.  

\textit{Diagrammatic calculus.}
By making use of the compact notation (\ref{G3l})-(\ref{G3c}), we can rewrite Eq. (\ref{RLS}) as
\begin{eqnarray}
\label{RLSc}
R^{(\mathrm{S})}_\mathsmaller{\Gamma_{1}}=\frac{4}{N}\frac{\mediaS{\Gamma_2}}{\mediaS{\Gamma_1}^2}-\frac{4}{N}+\frac{1}{N^2\mediaS{\Gamma_1}}.
\end{eqnarray}
We see that the main term is given by the ratio between a $\mediaS{\cdot}$-average of two links sharing a common node, and the square of 
the $\mediaS{\cdot}$-average of a single link,\textit{i.e.}, our motif. 
In fact, for any motif $\Gamma$, we can express $R^{(\mathrm{S})}_\mathsmaller{\Gamma}$ by using the diagrammatic notation. 
Furthermore, since $R^{(\mathrm{S})}_\mathsmaller{\Gamma}$ is defined in terms
of connected correlation functions (\ref{RNS}), we need to work only with Green functions of connected motifs, as the contributions
of disconnected motifs always cancel. 
Next, by using this diagrammatic tool, we evaluate $R^{(\mathrm{S})}_\mathsmaller{\Gamma}$ in the crucial case of $k$-cliques.
We first analyze the case $k=3$ in detail, then we look at the general behavior $R^{(\mathrm{S})}_\mathsmaller{\Gamma_{kc}}$,
and finally we extrapolate the general behavior for  any motif.

\begin{figure}[tbh]
{\includegraphics[height=3.5in]{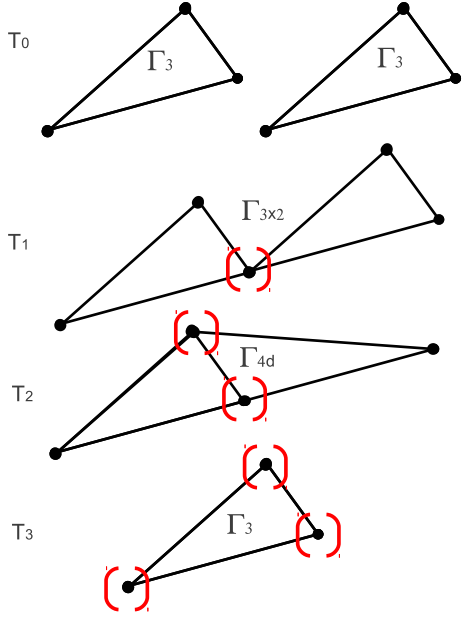}}
%{\includegraphics[height=3.5in]{figures/fig10_PRE.pdf}}
\caption{Contributions to Eq. (\ref{RTS2}) (panels $\mathrm{T}_0$-$\mathrm{T}_3$), and to Eq. (\ref{RTS3}) (panels $\mathrm{T}_1$-$\mathrm{T}_3$).
Two randomly chosen triangles in the network may have from zero to full overlap. The red parenthesis indicate the shared nodes.
$\mathrm{T}_0$ represents the case with two disconnected triangles, $\mathrm{T}_1$ represents two triangles sharing a common node, $\mathrm{T}_2$ represents two triangles sharing 2 nodes
(and then one link), $\mathrm{T}_3$ two triangles sharing 3 nodes (and then totally overlapped). As stressed before, in calculating
the connected correlation functions (\ref{RNS}-\ref{RNQB}), two disconnected motifs ($\mathrm{T}_0$ in the present case) do not contribute.  
\label{fig10}
}
\end{figure}

\textit{Triangle} $(\Gamma_{3})$. 
This is the simplest $k$-clique. Now $k_{\mathsmaller{\Gamma_{3}}}(i)$ counts the number of triangles passing through the node $i$:
\begin{eqnarray}
\label{RTS0}
2 ~ k_{\mathsmaller{\Gamma_{3}}}(i) =\sum_{l,m: l\neq m}a_{i,l}a_{i,m}a_{l,m},
\end{eqnarray}
where $c=3$ and the factor 2 comes from the fact that in the rhs we have not restricted the sum to ordered couples $l<m$.
However, we stress again that these constants do not play any role for the standard ratios. 
From Eq. (\ref{RTS0}) we have (notice that $a_{i,i}=0$, but $p(h_i,h_i)\neq 0$)
%
%\begin{eqnarray}
%\label{RTS1}
%2\mediaT{k_{\mathsmaller{\Gamma_{3}}}(i)}=\left(N^2-3N\right)\mediaS{p(h_1,h_2)p(h_2,h_3)p(h_3,h_1)},
%\end{eqnarray}
%\begin{eqnarray}
%\label{RTS2}
%&& (2)^2\mediaT{k_{\mathsmaller{\Gamma_{3}}}(i)k_{\mathsmaller{\Gamma_{3}}}(j)}=\nonumber \\
%&& \left(N^4-13N^3\right)\mediaS{p(h_1,h_2)p(h_2,h_3)p(h_3,h_1)}^2\nonumber \\
%&& +8N^3\mediaS{p(h_1,h_5)p(h_2,h_5)p(h_1,h_2)p(h_3,h_5)p(h_4,h_5)p(h_3,h_4)}\nonumber \\
%&& +10N^2\mediaS{p(h_1,h_3)p(h_1,h_4)p(h_2,h_3)p(h_2,h_4)}\nonumber \\
%&& +4N\mediaS{p(h_1,h_2)p(h_2,h_3)p(h_3,h_1)},
%\end{eqnarray}
%\begin{eqnarray}
%\label{RTS3}
%&& (2)^2\mediaT{k^2_{\mathsmaller{\Gamma_{3}}}(i)}=\nonumber \\
%&& \left(N^4-10N^3\right)\mediaS{p(h_1,h_2)p(h_2,h_3)p(h_3,h_1)}^2\nonumber \\
%&& +4N^3\mediaS{p(h_1,h_3)p(h_1,h_4)p(h_2,h_3)p(h_2,h_4)}+\nonumber \\
%&& 4N^2\mediaS{p(h_1,h_2)p(h_2,h_3)p(h_3,h_1)}.
%\end{eqnarray}
%
\begin{eqnarray}
\label{RTS1}
2\mediaT{k_{\mathsmaller{\Gamma_{3}}}(i)}=\left(N-1\right)\left(N-2\right)\mediaS{\Gamma_{3}},
\end{eqnarray}
and upon referring to the contributions as in Fig. (\ref{fig10}), 
and taking into account the multiplicities of all the topologically equivalent cases, we get
\begin{eqnarray}
\label{RTS2}
&& (2)^2\mediaT{k_{\mathsmaller{\Gamma_{3}}}(i)k_{\mathsmaller{\Gamma_{3}}}(j)}=\nonumber \\
&& \left(N-2\right)\left(N-3\right)\left(N-4\right)\left(N-5\right)\mediaS{\Gamma_{3}}^2\nonumber \\
&& +8\left(N-2\right)\left(N-3\right)\left(N-4\right)\mediaS{\Gamma_{3\times 2}}\nonumber \\
&& +10\left(N-2\right)\left(N-3\right)\mediaS{\Gamma_{4 d}}
+4\left(N-2\right)\mediaS{\Gamma_{3}},
\end{eqnarray}
\begin{eqnarray}
\label{RTS3}
&& (2)^2\mediaT{k^2_{\mathsmaller{\Gamma_{3}}}(i)}=
\left(N-1\right)\left(N-2\right)\left(N-3\right)\left(N-4\right)\mediaS{\Gamma_{3}}^2\nonumber \\
&& +4\left(N-1\right)\left(N-2\right)\left(N-3\right)\mediaS{\Gamma_{4d}}\nonumber \\
&& + 2\left(N-1\right)\left(N-2\right)\mediaS{\Gamma_{3}}.
\end{eqnarray}
By keeping for each term only the leading and subleading contributions we get
\begin{eqnarray}
\label{RTS4}
2\mediaT{k_{\mathsmaller{\Gamma_{3}}}(i)}=\left(N^2-3N\right)\mediaS{\Gamma_{3}},
\end{eqnarray}
\begin{eqnarray}
\label{RTS5}
&& (2)^2\mediaT{k_{\mathsmaller{\Gamma_{3}}}(i)k_{\mathsmaller{\Gamma_{3}}}(j)}=\left(N^4-13N^3\right)\mediaS{\Gamma_{3}}^2\nonumber \\
&& +8N^3\mediaS{\Gamma_{3\times 2}}
+10N^2\mediaS{\Gamma_{4 d}}
+4N\mediaS{\Gamma_{3}},
\end{eqnarray}
\begin{eqnarray}
\label{RTS6}
&& (2)^2\mediaT{k^2_{\mathsmaller{\Gamma_{3}}}(i)}=\left(N^4-10N^3\right)\mediaS{\Gamma_{3\times 2}}\nonumber \\
&& +4N^3\mediaS{\Gamma_{4d}}+ 4N^2\mediaS{\Gamma_{3}}.
\end{eqnarray}
Plugging Eqs. (\ref{RTS4}-\ref{RTS6}) into Eq. (\ref{RS}) via Eq. (\ref{RNS}) we have finally
\begin{eqnarray}
\label{RTS}
%R_\mathsmaller{\Gamma_{3}}=\frac{9}{N}\frac{\mediaS{\Gamma_{3\times 2}}}{\mediaS{\Gamma_{3}}}-\frac{9}{N}+\mathrm{O}\left(\frac{1}{N}\right).
R^{(\mathrm{S})}_\mathsmaller{\Gamma_{3}}=\frac{9}{N}\frac{\mediaS{\Gamma_{3\times 2}}}{\mediaS{\Gamma_{3}}^2}-\frac{9}{N}+\frac{14}{N^2}\frac{\mediaS{\Gamma_{4d}}}{\mediaS{\Gamma_{3}}^2}
+\frac{6}{N^3}\frac{1}{\mediaS{\Gamma_{3}}}.
\end{eqnarray}
Similarly to the last term of Eq. (\ref{RLS}),
the last two terms of Eq. (\ref{RTS}) do not cancel for large $\gamma$ (see Figs. \ref{fig14} and \ref{fig15}).  
%Here the most important contribution comes from the self-term $i=j$ and it is contained in $\mediaS{\Gamma_{4c}}$.

%\textit{$4$-Clique} $(\Gamma_{4c})$. If $\Gamma$ is the $4$-clique, we have
%\begin{eqnarray}
%\label{RQd}
%R_\mathsmaller{\Gamma_{4c}}=\frac{16}{N}\frac{\mediaS{\Gamma_{4f\times 2}}}{\mediaS{\Gamma_{4c}}^2}-\frac{16}{N}+\mathrm{O}\left(\frac{1}{N}\right),
%\end{eqnarray}
%where the last term palys a role similar to the last term of Eq. (\ref{RT}), while the first can diverge. 
%The factor 16 is due to the symmetry of the motif $\Gamma_{4c}$, we have in fact 15 ways to build $\Gamma_{4f\times 2}$ from the correlated terms with $i\neq j$, and
%an extra contribution from the self-term $i=j$.

\textit{$k$-Clique $(\Gamma_{kc})$ and the general case.} 
We are able to write down a general formula for the general motif.
Later on, we will show that, given the motif $\Gamma$, the most important and interesting contributions to the fluctuations are those corresponding 
to the cases in which two randomly chosen motifs share a single node. So, for the triangle case ($\Gamma=\Gamma_3$), this corresponds to panel $\mathrm{T}_1$ in Fig. (\ref{fig10}). 
In particular, it is easy to derive the main contribution when the motif is a $k$-\textit{clique},
$(\Gamma_{kc})$. In fact, given ${\Gamma_{kc}}$ ($(k-1)$ is the degree of each node), if $\Gamma_{kc\times 2}$ indicates the motif
in which two $k$-cliques $\Gamma_{kc}$ share a common node, we have  
\begin{eqnarray}
\label{Rclique}
R^{(\mathrm{S})}_\mathsmaller{\Gamma_{kc}}=\frac{b_k}{N}\frac{\mediaS{\Gamma_{kc\times 2}}}{\mediaS{\Gamma_{kc}}^2}-\frac{b_k}{N}+\mathrm{O}\left(\frac{1}{N}\right),
\end{eqnarray}
where $b_k$ is a combinatorial term which depends only on $k$, and the last term is positive and plays a role similar to the last two terms of Eq. (\ref{RTS}).

More in general, if the motif $\Gamma$ is a star, or any not regular motif (regular polygons and $k$-\textit{cliques} are example of regular motifs),
the most important contribution to $R^{(\mathrm{S})}_\mathsmaller{\Gamma}$ corresponds to the case in which two randomly chosen $\Gamma$'s share
the node having the maximal degree of $\Gamma$, $k_{\max}$. If we indicate by $\Gamma_{k_{\max}\times 2}$ this motif, we have
\begin{eqnarray}
\label{RGEN}
R^{(\mathrm{S})}_\mathsmaller{\Gamma}=\frac{b_\mathsmaller{\Gamma}}{N}\frac{\mediaS{\Gamma_{k_{\max}\times 2}}}{\mediaS{\Gamma}^2}-\frac{b_\mathsmaller{\Gamma}}{N}+\ldots
\end{eqnarray}

\subsection{$R^{(\mathrm{A})}$ (Sampling the hidden-variables by first freezing the $h$'s)}
Here we consider to sampling by first freezing the node-hidden-variables, and only at the end averaging with respect to them.
We have then to use Eq. (\ref{RNA}). As we will soon see, the calculation is quite close to the symmetric (S) case, as well
as the behavior for finite $\gamma$, while there emerges an important difference for the classical random graph limit in which $\gamma\gg 1$.
Let us first calculate $R^{(\mathrm{A})}$ in a few crucial cases. 

\textit{Link} $(\Gamma_{1})$.
The definition of $k_{\mathsmaller{\Gamma}}(i)$ has been given in Eq. (\ref{link}).
As stressed before, the averages over the two sets of hidden-variables do not change by changing their order.
Let us consider the product $k_{\mathsmaller{\Gamma_{1}}}(i)k_{\mathsmaller{\Gamma_{1}}}(j)$.
The crucial mathematical difference with respect to the symmetric case is that, in the calculation, 
we are now not facing the fact that $a^2_{i,j}=a_{i,j}$; rather, we will face the evaluation of things like $\mediaT{a_{i,j}}_{\bm{h}}\mediaT{a_{i,j}}_{\bm{h}}=p^2(h_i,h_j)$.
As before, we have to distinguish the case $i=j$ and $i\neq j$, see Fig.~\ref{fig9}. 
For $i=j$ we have
\begin{eqnarray}
\label{link3A}
&&\mediaS{\mediaT{k_{\mathsmaller{\Gamma_{1}}}(i)}^2_{|\bm{h}}}%=\sum_{l,m\neq i}\mediaS{p(h_i,h_l)p(h_i,h_m)}+\sum_{l\neq i}\mediaS{p(h_i,h_l)} \nonumber\\ && 
=(N-1)(N-2)\mediaS{p(h_1,h_2)p(h_2,h_3)}
\nonumber\\ && + (N-1)\mediaS{p^2(h_1,h_2)},
\end{eqnarray}
while for $i\neq j$ we have
\begin{eqnarray}
\label{link4A}
&& \mediaS{\mediaT{k_{\mathsmaller{\Gamma_{1}}}(i)}_{|\bm{h}}\mediaT{k_{\mathsmaller{\Gamma_{1}}}(j)}_{|\bm{h}}}%=\sum_{l\neq i,m\neq j:l\neq m}\mediaS{p(h_i,h_l)p(h_j,h_m)} 
%\nonumber\\ && + 3 \sum_{l\neq i,j}\mediaS{p(h_i,h_l)p(h_l,h_j)}=
=(N-2)(N-3)\mediaS{p(h_1,h_2)}^2
\nonumber\\ && + 3(N-2)\mediaS{p(h_1,h_2)p(h_2,h_3)}.
\end{eqnarray}
Comparing with the symmetric case, we see that, whereas Eqs. (\ref{link4}) and (\ref{link4A}) are identical,
Eqs. (\ref{link3}) and (\ref{link3A}) differ in the last term which here, due to the fact that the network is sparse,
is always smaller and negligible with respect to the other terms. 
Upon plugging Eq. (\ref{link1}) and Eqs. (\ref{link3A}-\ref{link4A}) into Eq. (\ref{RA}) via Eq. (\ref{RNA}) and, 
by keeping in this latter only terms in $N^2$, which cancel exactly, and terms in $N$, we obtain
\begin{eqnarray}
\label{RLA}
R^{(\mathrm{A})}_\mathsmaller{\Gamma_{1}}=\frac{4}{N}\frac{\mediaS{p(h_1,h_2)p(h_2,h_3)}}{\mediaS{p(h_1,h_2)}^2}-\frac{4}{N}.
\end{eqnarray}
With respect to the symmetric case, we observe now the absence of the last term of Eq. (\ref{RLS}).
As a consequence, the density of the node-degree in the A case, besides to be self-averaging,
has also the property that in the classical random graph limit, the relative fluctuations tend to zero:
\begin{eqnarray}
\label{RLAl}
\lim_{\gamma\gg 1} R^{(\mathrm{A})}_\mathsmaller{\Gamma_{1}}=0.
\end{eqnarray} 

\begin{figure}[tbh]
{\includegraphics[width=2.5in]{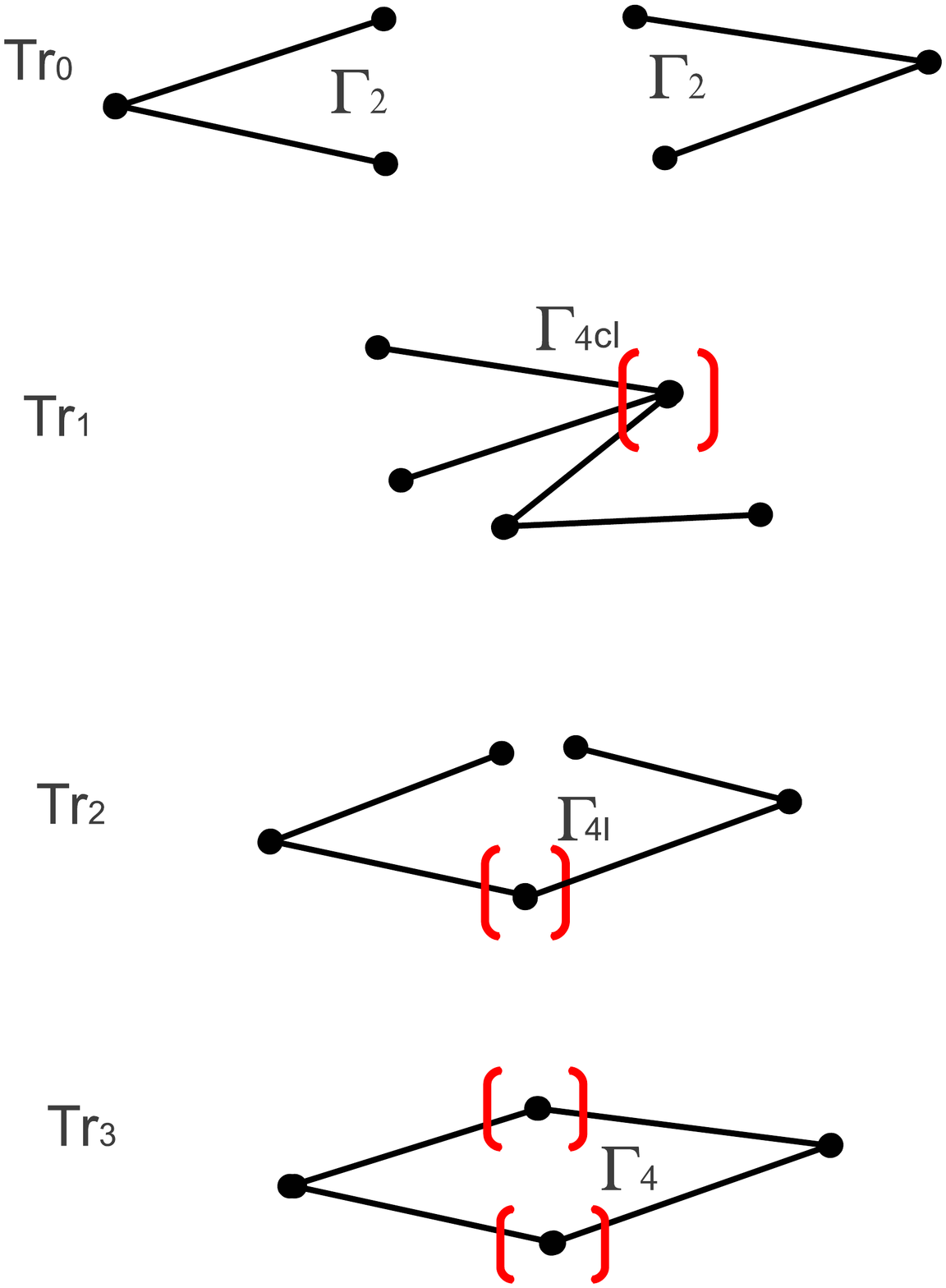}}
%{\includegraphics[width=2.5in]{figures/fig11_PRE.pdf}}
{\includegraphics[width=2.5in]{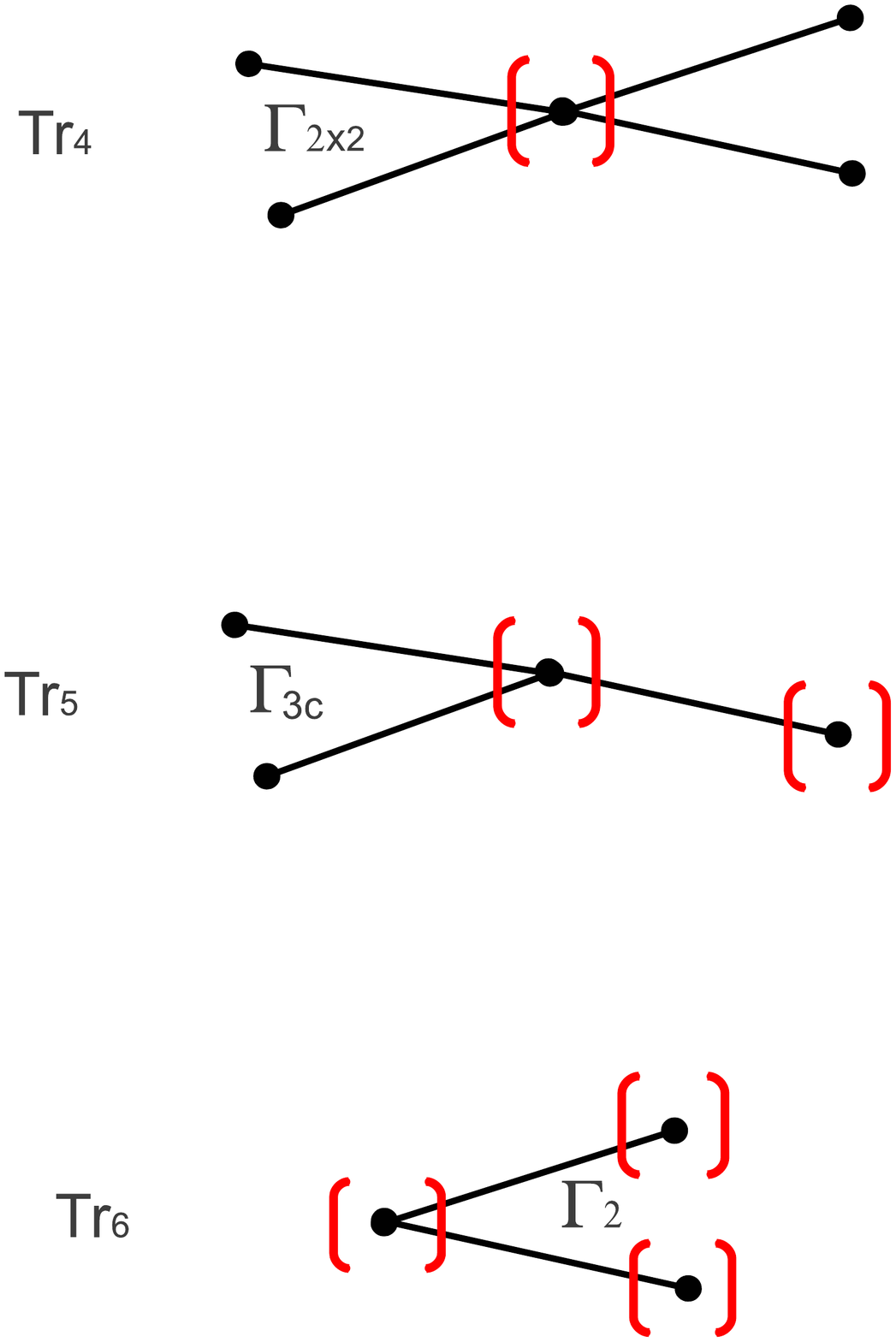}}
%{\includegraphics[width=2.5in]{figures/fig12_PRE.pdf}}
\caption{Contributions to Eq. (\ref{RTRA2}) (panels $\mathrm{Tr}_0$-$\mathrm{Tr}_3$), and to Eq. (\ref{RTRA3}) (panels $\mathrm{Tr}_4$-$\mathrm{Tr}_6$).
\label{fig11&12}
}
\end{figure}

\textit{Triple} $(\Gamma_{2})$.
Now $k_{\mathsmaller{\Gamma_{2}}}(i)$ counts the number of triples passing through the node $i$:
\begin{eqnarray}
\label{RTRA0}
2 ~ k_{\mathsmaller{\Gamma_{2}}}(i) =\sum_{l,m: l\neq m}a_{i,l}a_{i,m},
\end{eqnarray}
where $c=1$, since we count two consecutive links as a triple passing through node $i$ only if the center of the triple coincides with $i$;
and the factor 2 comes from the fact that in the rhs we have not restricted the sum to ordered couples $l<m$
(again, these constants do not play any role for the standard ratios). 
With respect to the triangle case, the triple case leads to a more involved calculation due to the lack of symmetry.
From Eq. (\ref{RTRA0}) we have (notice that $a_{i,i}=0$, but $p(h_i,h_i)\neq 0$)
\begin{eqnarray}
\label{RTRA1}
2\mediaT{k_{\mathsmaller{\Gamma_{2}}}(i)}=\left(N-1\right)\left(N-2\right)\mediaS{\Gamma_{2}}.
\end{eqnarray}
Upon referring to the contributions as in panels $\mathrm{Tr}_0$-$\mathrm{Tr}_3$ of Fig. (\ref{fig11&12}), 
and taking into account the multiplicities of all the topologically equivalent cases, we get
\begin{eqnarray}
\label{RTRA2}
&& (2)^2\mediaS{\mediaT{k_{\mathsmaller{\Gamma_{2}}}(i)}_{|\bm{h}}\mediaT{k_{\mathsmaller{\Gamma_{2}}}(j)}_{|\bm{h}}}=\nonumber \\
&& \left(N-2\right)\left(N-3\right)\left(N-4\right)\left(N-5\right)\mediaS{\Gamma_{2}}^2\nonumber \\
&& +4\left(N-2\right)\left(N-3\right)\left(N-4\right)\mediaS{\Gamma_{4\mathrm{l}}}\nonumber \\
&& +4\left(N-2\right)\left(N-3\right)\left(N-4\right)\mediaS{\Gamma_{4\mathrm{cl}}}\nonumber \\
&& +2\left(N-2\right)\left(N-3\right)\mediaS{\Gamma_{4}},
\end{eqnarray}
and upon referring to panels $\mathrm{Tr}_4$-$\mathrm{Tr}_6$ of Fig. (\ref{fig11&12})
\begin{eqnarray}
\label{RTRA3}
&& (2)^2\mediaS{\mediaT{k_{\mathsmaller{\Gamma_{2}}}(i)}_{|\bm{h}}^2}= \nonumber \\
&& \left(N-1\right)\left(N-2\right)\left(N-3\right)\left(N-4\right)\mediaS{\Gamma_{2\times 2}}\nonumber \\
&& +5\left(N-1\right)\left(N-2\right)\left(N-3\right)\mediaS{\Gamma_{3\mathrm{c}}}\nonumber \\
&& +3\left(N-1\right)\left(N-2\right)\mediaS{\Gamma_{2}}.
\end{eqnarray}

Finally, plugging Eq. (\ref{RTRA1}) and Eqs. (\ref{RTRA2}-\ref{RTRA3}) into Eq. (\ref{RA}) via Eq. (\ref{RNA}) and, 
by keeping in this latter only terms in $N^2$, which cancel exactly, and terms in $N$, we obtain
\begin{eqnarray}
\label{RTRA}
R^{(\mathrm{A})}_\mathsmaller{\Gamma_{2}}=\frac{1}{N}\frac{4\mediaS{\Gamma_{4\mathrm{l}}}+4\mediaS{\Gamma_{4\mathrm{cl}}}+\mediaS{\Gamma_{2\times 2} }}{\mediaS{\Gamma_{2}}^2}-\frac{9}{N}.
\end{eqnarray}
Again, we observe that
\begin{eqnarray}
\label{RLAl}
\lim_{\gamma\gg 1} R^{(\mathrm{A})}_\mathsmaller{\Gamma_{2}}=0.
\end{eqnarray} 

\begin{figure}[tbh]
{\includegraphics[width=2.5in]{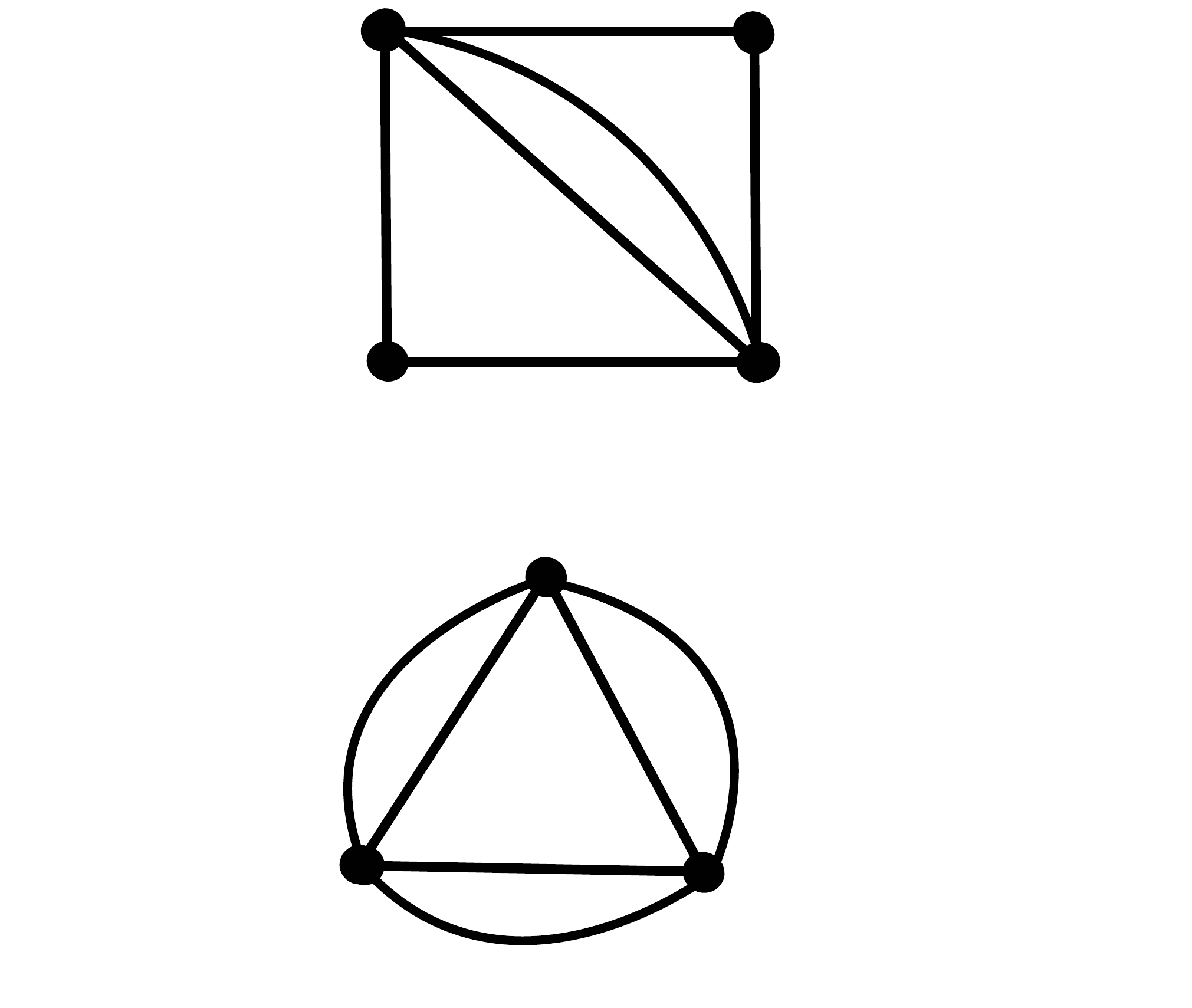}}
%{\includegraphics[width=2.5in]{figures/fig13_PRE.pdf}}
\caption{Negligible contributions to Eq. (\ref{RTA5}) (upper), and to Eq. (\ref{RTA6}) (lower).
Notice that these motifs having multiple links are only a useful artifact of the calculation, 
but there are no motifs with multiple links in the graph model (the hidden-variable model is by construction a \textit{simple graph}, 
\textit{i.e.}, a graph with no multiple links or loop-links). 
\label{fig13}
}
\end{figure}

\textit{Triangle} $(\Gamma_{3})$.
The calculation for the triangle case is much easier due to the full symmetry of the motif,
and it runs in close analogy to the calculation already done in the S case (the symmetric sampling). 
In fact, instead of Eqs. (\ref{RTS5}-\ref{RTS6}), now we have 
\begin{eqnarray}
\label{RTA5}
&& (2)^2\mediaS{\mediaT{k_{\mathsmaller{\Gamma_{3}}}(i)}_{|\bm{h}}\mediaT{k_{\mathsmaller{\Gamma_{3}}}(j)}_{|\bm{h}}}=\left(N^4-13N^3\right)\mediaS{\Gamma_{3}}^2\nonumber \\
&& +8N^3\mediaS{\Gamma_{3\times 2}}
+10N^2\mediaS{\Gamma_{4 d}p},
\end{eqnarray}
\begin{eqnarray}
\label{RTA6}
&& (2)^2\mediaS{\mediaT{k_{\mathsmaller{\Gamma_{3}}}(i)}^2_{|\bm{h}}}=\left(N^4-10N^3\right)\mediaS{\Gamma_{3\times 2}}\nonumber \\
&& +4N^3\mediaS{\Gamma_{4d}p}+ 4N^2\mediaS{\Gamma_{3}ppp},
\end{eqnarray}
where, the short notations $\Gamma_{4d}p$ and $\Gamma_{3}ppp$ corresponds to the motifs with multiple links shown in Fig. \ref{fig13}.

In fact, for large $N$, the contributions of these motifs are always negligible with respect to the others, and for the standard ratio we obtain
\begin{eqnarray}
\label{RTA}
R^{(\mathrm{A})}_\mathsmaller{\Gamma_{3}}=\frac{9}{N}\frac{\mediaS{\Gamma_{3\times 2}}}{\mediaS{\Gamma_{3}}^2}-\frac{9}{N},
\end{eqnarray}
so that, again, unlike Eq. (\ref{RTS}), we have
\begin{eqnarray}
\label{RTAl}
\lim_{\gamma\gg 1} R^{(\mathrm{A})}_\mathsmaller{\Gamma_{3}}=0,
\end{eqnarray} 
and for the general $k$-cliques we have
\begin{eqnarray}
\label{RcliqueA}
R^{(\mathrm{A})}_\mathsmaller{\Gamma_{kc}}=\frac{b_k}{N}\frac{\mediaS{\Gamma_{kc\times 2}}}{\mediaS{\Gamma_{kc}}^2}-\frac{b_k}{N},
\end{eqnarray}
where $b_k$ is the same combinatorial factor which appears in Eq. (\ref{Rclique}).

\subsection{$R^{(\mathrm{B})}$ (Sampling the hidden-variables by first freezing the $q$'s)}
Here we consider to sampling by first freezing the link-hidden-variables, and only at the end averaging with respect to them.
We have then to use Eq. (\ref{RNB}), and make use of the $\{\}$-averages, defined by Eq. (\ref{aveq}).
The diagrammatic calculus now has very different rules.
Let us first calculate $R^{(\mathrm{B})}$ in a few crucial cases. 

\textit{Link} $(\Gamma_{1})$.
The definition of $k_{\mathsmaller{\Gamma_1}}(i)$ has been given in Eq. (\ref{link}).
Due to Eq. (\ref{Equi}), we already know that it holds the equivalence of the ensembles (S),  (A) and (B) for their averages.
In fact, the global average requires averaging over two independent set of hidden-variables.
It is however instructive to show here the details of the two processes.
For the quenched averages of the adjacency matrix we have 
\begin{eqnarray}
\label{linkB}
\mediaT{a_{i,j}}_{|\bm{q}}=\mediaS{\theta\left(p(h_i,h_j)-q_{i,j}\right)},
\end{eqnarray}
where $i$ and $j$ are any two different indices, and
where $\theta(\cdot)$ is the Heaviside step function ($\theta(u)=1$, for $u\geq 0$, and 0 otherwise).
From Eq. (\ref{linkB}), by using the property
\begin{eqnarray}
\label{linkB1}
\int_0^1 \theta(u-q)dq=u,
\end{eqnarray}
valid for $0\leq u\leq 1$, it is immediate to check that Eq. (\ref{link}) leads to
\begin{eqnarray}
\label{linkB2}
\left\{\mediaT{k_\mathsmaller{\Gamma_{1}}(i)}_{|\bm{q}}\right\}=(N-1)\left\{\mediaT{a_{i,j}}_{|\bm{q}}\right\}=\mediaS{p(h,h')}.
\end{eqnarray}
Let us now consider the product $k_{\mathsmaller{\Gamma_{1}}}(i)k_{\mathsmaller{\Gamma_{1}}}(j)$.
For $i=j$ we have
\begin{eqnarray}
\label{linkB4}
&&\mediaT{k_{\mathsmaller{\Gamma_{1}}}(i)}^2_{|\bm{q}}=\nonumber\\
&& \sum_{l,m: l,m\neq i} \mediaS{\theta\left(p(h_i,h_l)-q_{i,l}\right)}\mediaS{\theta\left(p(h_i,h_m)-q_{i,m}\right)}.~~~~~
\end{eqnarray}
Now, in the above sum we have $N-1$ non trivial correlated terms corresponding to the cases $l=m$.
In fact, by taking also the average over the $q$'s (which are random variables uniformly distributed over the interval $[0,1]$),
we have
\begin{eqnarray}
\label{linkB5}
&&\left\{\mediaT{k_\mathsmaller{\Gamma_{1}}(i)}^2_{|\bm{q}}\right\}=(N-1)(N-2)\mediaS{p(h,h')}^2+(N-1)\nonumber\\
&&\times \int_0^1 dq \mediaS{\theta\left(p(h_1,h_2)-q\right)}\mediaS{\theta\left(p(h_3,h_4)-q\right)}.
\end{eqnarray}
By integrating the $\theta$ function we get the non trivial term 
\begin{eqnarray}
\label{linkB6}
&&\int_0^1 dq \mediaS{\theta\left(p(h_1,h_2)-q\right)}\mediaS{\theta\left(p(h_3,h_4)-q\right)}\nonumber \\
&&=\mediaS{\min\{p(h_1,h_2);p(h_3,h_4)\}}.
\end{eqnarray}

For $i\neq j$ we have
\begin{eqnarray}
\label{linkB3}
&& \mediaT{k_{\mathsmaller{\Gamma_{1}}}(i)}_{|\bm{q}}\mediaT{k_{\mathsmaller{\Gamma_{1}}}(j)}_{|\bm{q}}=\nonumber\\
&&\sum_{l,m: l\neq i, m\neq j} \mediaS{\theta\left(p(h_i,h_l)-q_{i,l}\right)}\mediaS{\theta\left(p(h_j,h_m)-q_{j,m}\right)}.~~~~~
\end{eqnarray}
We observe now that, in the above sum, unlike the (S) and (A) ensembles, for almost all pairs of links, $(i,l)$ and $(j,m)$, 
the random variables $q_{i,l}$ and $q_{j,m}$ are independent.
In fact, by referring to Fig. \ref{fig9} with $i\neq j$, none of the shown diagrams contribute to the connected correlation function.
More precisely, in the sum of Eq. (\ref{linkB3}), there is only one term which is non trivial and corresponds to the case
$l=j,m=i$ and is given by Eq. (\ref{linkB6}).
Note that in the (S) and (A) ensembles, for $N$ large enough, such link-overlap contributions were always negligible with respect
to the node-sharing contributions.  
In conclusion, from the above Eqs. and observations, for $N$ large enough, we get
\begin{eqnarray}
\label{RLB}
R^{(\mathrm{B})}_\mathsmaller{\Gamma_{1}}=\frac{2}{N}\frac{\mediaS{\min\{p(h_1,h_2);p(h_3,h_4)\}}}{\mediaS{p(h_1,h_2)}^2}.
\end{eqnarray}
From Eq. (\ref{RLB} ), and observing that 
\begin{eqnarray}
\label{linkB7}
\mediaS{\min\{p(h_1,h_2);p(h_3,h_4)\}}<\mediaS{p(h,h')},
\end{eqnarray}
we conclude that for large $N$ we obtain again $R^{(\mathrm{B})}=\mathrm{O}(1/N)$.  
However, with respect to the (S) and (A) cases, we observe a quite different structure of $R^{(\mathrm{B})}$.
In the next example we will see that these differences play a crucial role for larger motifs. 

\textit{Triangle} $(\Gamma_{3})$. 
From Eq. (\ref{RTS0}) we have
\begin{eqnarray}
\label{RTB0}
&& 2 ~ \mediaT{k_{\mathsmaller{\Gamma_{3}}}(i)}_{|\bm{q}}=\sum_{l,m: l\neq m; l,m\neq i}\mediaS{\theta\left(p(h_i,h_l)-q_{i,l}\right)\right. \nonumber \\
&&\times \left. \theta\left(p(h_l,h_m)-q_{l,m}\right)\theta\left(p(h_i,h_m)-q_{i,m}\right)}.
\end{eqnarray}
Therefore we need to evaluate
\begin{eqnarray}
\label{RTB1}
&& (2)^2 ~ \mediaT{k_{\mathsmaller{\Gamma_{3}}}(i)}_{|\bm{q}}\mediaT{k_{\mathsmaller{\Gamma_{3}}}(j)}_{|\bm{q}}=\nonumber \\
&& \sum_{l,m,l',m': l\neq m, l'\neq m'; l,m\neq i;l',m'\neq j}
\mediaS{\theta\left(p(h_i,h_l)-q_{i,l}\right)\right. \nonumber \\
&&\times \left. \theta\left(p(h_l,h_m)-q_{l,m}\right)\theta\left(p(h_i,h_m)-q_{i,m}\right)}\nonumber \\
&&\times
\mediaS{\theta\left(p(h_j,h_{l'})-q_{j,l'}\right)\right. \nonumber \\
&&\times \left. \theta\left(p(h_{l'},h_{m'})-q_{l',m'}\right)\theta\left(p(h_j,h_{m'})-q_{j,m'}\right)}.
\end{eqnarray}
and similarly
\begin{eqnarray}
\label{RTB2}
&& (2)^2 ~ \mediaT{k_{\mathsmaller{\Gamma_{3}}}(i)}^2_{|\bm{q}}=\nonumber \\
&& \sum_{l,m,l',m': l\neq m, l'\neq m'; l,m,l',m'\neq i}
\mediaS{\theta\left(p(h_i,h_l)-q_{i,l}\right)\right. \nonumber \\
&&\times \left. \theta\left(p(h_l,h_m)-q_{l,m}\right)\theta\left(p(h_i,h_m)-q_{i,m}\right)}\nonumber \\
&&\times
\mediaS{\theta\left(p(h_i,h_{l'})-q_{i,l'}\right)\right. \nonumber \\
&&\times \left. \theta\left(p(h_{l'},h_{m'})-q_{l',m'}\right)\theta\left(p(h_i,h_{m'})-q_{i,m'}\right)}.
\end{eqnarray}
When we average Eqs. (\ref{RTB1}) and (\ref{RTB2}) over the $q$'s, we have the same topological combinations
of Fig. (\ref{fig10}).
However, each diagrammatic contribution is now different with respect to the
(S) and (A) cases. In fact, upon referring to the contributions as in Fig. (\ref{fig10}) we have
\begin{eqnarray}
\label{RTB1b}
&& (2)^2\mediaG{\mediaT{k_{\mathsmaller{\Gamma_{2}}}(i)}_{|\bm{q}}\mediaT{k_{\mathsmaller{\Gamma_{2}}}(j)}_{|\bm{q}}}=\nonumber \\
&& \left(N-2\right)\left(N-3\right)\left(N-4\right)\left(N-5\right)\mediaS{\Gamma_{3}}^2\nonumber \\
&& +8\left(N-2\right)\left(N-3\right)\left(N-4\right)\mediaS{\Gamma_{3}}^2\nonumber \\
&& +10\left(N-2\right)\left(N-3\right)\left(N-4\right)\mediaS{\widetilde{\Gamma_{4\mathrm{d}}}}\nonumber \\
&& +2\left(N-2\right)\left(N-3\right)\mediaS{\widetilde{\Gamma_{3}}},
\end{eqnarray}
and 
\begin{eqnarray}
\label{RTB2b}
&& (2)^2\mediaG{\mediaT{k_{\mathsmaller{\Gamma_{2}}}(i)}_{|\bm{q}}^2}= \nonumber \\
&& \left(N-1\right)\left(N-2\right)\left(N-3\right)\left(N-4\right)\mediaS{\Gamma_{3}}^2\nonumber \\
&& +4\left(N-1\right)\left(N-2\right)\left(N-3\right)\mediaS{\widetilde{\Gamma_{4\mathrm{d}}}}\nonumber \\
&& +2\left(N-1\right)\left(N-2\right)\mediaS{\widetilde{\Gamma_{3}}},
\end{eqnarray}
where $\mediaS{\widetilde{\Gamma_{4\mathrm{d}}}}$ and $\mediaS{\widetilde{\Gamma_{3}}}$ are a slight modification
of $\mediaS{\Gamma_{4\mathrm{d}}}$ and $\mediaS{\Gamma_{3}}$, respectively:
\begin{eqnarray}
\label{tilde1}
&& \mediaS{\widetilde{\Gamma_{4\mathrm{d}}}}=\mediaS{p(h_i,h_l)p(h_i,h_m)p(h_i',h_l')p(h_i',h_m') \right. \nonumber \\ 
&\times & \left. \min\left\{p(h_l,h_m);p(h_l',h_m')\right\} },
\end{eqnarray}
\begin{eqnarray}
\label{tilde2}
&& \mediaS{\widetilde{\Gamma_{3}}}=\mediaS{\min\left\{p(h_i,h_l);p(h_i',h_l')\right\}  \right. \nonumber \\ 
&\times &  \min\left\{p(h_i,h_m);p(h_i',h_m')\right\} \nonumber \\
&\times & \left. \min\left\{p(h_l,h_m);p(h_l',h_m')\right\} }.
\end{eqnarray}
From Eqs. (\ref{RTB1b}) and (\ref{RTB2b}) we get
\begin{eqnarray}
\label{RTB1c}
&& (2)^2\mediaG{\mediaT{k_{\mathsmaller{\Gamma_{2}}}(i)}_{|\bm{q}}\mediaT{k_{\mathsmaller{\Gamma_{2}}}(j)}_{|\bm{q}}}-
(2)^2\mediaG{\mediaT{k_{\mathsmaller{\Gamma_{2}}}(i)}_{|\bm{q}}}^2=\nonumber \\
&&\left(-14N^2+12N+4\right)\mediaS{\Gamma_{3}}^2+
10\left(N^3-9N^2\right)\mediaS{\widetilde{\Gamma_{4\mathrm{d}}}}\nonumber \\
&& + 2\left(N^2-5N\right)\mediaS{\widetilde{\Gamma_3}},
\end{eqnarray}
and
\begin{eqnarray}
\label{RTB2c}
&& (2)^2\mediaG{\mediaT{k_{\mathsmaller{\Gamma_{2}}}(i)}^2_{|\bm{q}}}-
(2)^2\mediaG{\mediaT{k_{\mathsmaller{\Gamma_{2}}}(i)}_{|\bm{q}}}^2=\nonumber \\
&&\left(-4N^3+22N^2\right)\mediaS{\Gamma_{3}}^2+
4N^3\mediaS{\widetilde{\Gamma_{4\mathrm{d}}}} \nonumber \\
&& + 2N^2\mediaS{\widetilde{\Gamma_3}}.
\end{eqnarray}
Taking into account that $\mediaS{\widetilde{\Gamma_{4\mathrm{d}}}}=\mathrm{O}(1/N^5)$, 
$\mediaS{\Gamma_3}=\mathrm{O}(1/N^3)$ and that $\mediaS{\widetilde{\Gamma_3}}<\mediaS{\Gamma_3}$,
from Eqs. (\ref{RB}) and (\ref{RNB}), for large $N$ we obtain
\begin{eqnarray}
\label{RTB}
R^{(\mathrm{B})}_\mathsmaller{\Gamma_3}=\frac{10 \mediaS{\widetilde{\Gamma_{4\mathrm{d}}}}}{N\mediaS{\Gamma_{3}}^2}+
\frac{2 \mediaS{\widetilde{\Gamma_3}}}{N^2\mediaS{\Gamma_3}^2}.
\end{eqnarray}
The two terms in the rhs of Eq. (\ref{RTB}) cannot generate singular terms for finite $\gamma$.
However, as we had anticipated, for $\gamma \gg 1$,  $R^{(\mathrm{B})}_\mathsmaller{\Gamma_3}$ become a growing function of $N$.

\subsection{Sampling only the $q$'s (\textit{i.e.}, frozen $h$'s)}
If the $h$'s remain frozen, it is interesting to evaluate $R^{(\mathrm{Q})}_{\mathsmaller{\Gamma}|\bm{h}}$
according to the definition (\ref{Rh}). When the $h$' s are frozen, we have to average only with respect to the $q$' s.
Now, if $i\neq j$, the mixed-term in Eq. (\ref{RNQA}) is actually identically null for each network realization:
\begin{eqnarray}
\label{Qh}
\mediaT{k_{\mathsmaller{\Gamma}}(i)k_{\mathsmaller{\Gamma}}(j)}_{i\neq j~|\bm{h}}-\mediaT{k_{\mathsmaller{\Gamma}}(i)}_{|\bm{h}}^2=0.
\end{eqnarray}
Eqs. (\ref{Rh}), (\ref{RNQA}) and (\ref{Qh}) imply
\begin{eqnarray}
\label{Rh1}
R^{(\mathrm{Q})}_{\mathsmaller{\Gamma}|\bm{h}}=\frac{\mediaT{k^2_{\mathsmaller{\Gamma}}(i)}_{|\bm{h}}-\mediaT{k_{\mathsmaller{\Gamma}}(i)}_{|\bm{h}}^2}
{N\mediaT{k_{\mathsmaller{\Gamma}}(i)}_{|\bm{h}}^2}.
\end{eqnarray}
Furthermore, since the $h$' are frozen, unlike the ensembles (S) and (A), 
the leading term in $\mediaT{k^2_{\mathsmaller{\Gamma}}(i)}_{|\bm{h}}$ always factorizes, so that, in the numerator of the rhs of Eq. (\ref{Rh1}) the leading term always cancels.
As a consequence, if for a motif $\Gamma$ we have $\mediaT{k_{\mathsmaller{\Gamma}}(i)}_{|\bm{h}}=\mathrm{O}(N^m p^n)$, where the integers $m$ and $n$
characterize the structure of $\Gamma$, we have $\mediaT{k^2_{\mathsmaller{\Gamma}}(i)}_{|\bm{h}}-\mediaT{k_{\mathsmaller{\Gamma}}(i)}^2_{|\bm{h}}=\mathrm{O}(N^{2m-1} p^{2n})$,
so that it is always 
\begin{eqnarray}
\label{Rh2}
R^{(\mathrm{Q})}_{\mathsmaller{\Gamma}|\bm{h}}=\mathrm{O}\left(\frac{1}{N}\right).
\end{eqnarray}
It is not difficult to verify the general Eq. (\ref{Rh2}) in a few cases.
For example, in the case in which $\Gamma$ is the link, we easily arrive at
\begin{eqnarray}
\label{Rh3}
R^{(\mathrm{Q})}_{\mathsmaller{\Gamma}|\bm{h}}=\frac{\sum_{i,l\neq i}\left(p(h_i,h_l)-p^2(h_i,h_l)\right)}
{\sum_{i,j,l,m:l,m\neq i}\left(p(h_i,h_l)p(h_i,h_m)\right)}=\mathrm{O}\left(\frac{1}{N}\right).
\end{eqnarray}
Similar arguments apply for the cases in which $\Gamma=\Gamma_2$ or $\Gamma=\Gamma_3$.
Of course, the exact value of $R^{(\mathrm{Q})}_{\mathsmaller{\Gamma}|\bm{h}}$ depends on the actual sequence $\{h_i\}$,
but for $N$ sufficiently large Eq. (\ref{Rh2}) always applies. 

\subsection{Sampling only the $h$'s (\textit{i.e.}, frozen $q$'s)}
If the $q$'s remain frozen, it is interesting to evaluate $R^{(\mathrm{Q})}_{\mathsmaller{\Gamma}|\bm{q}}$
according to the definition (\ref{Rq}). When the $q$' s are frozen, we have to average only with respect to the $h$' s.
It is not necessary to perform a detailed analysis to recognize what is the general behavior of $R^{(\mathrm{Q})}_{\mathsmaller{\Gamma}|\bm{q}}$.
In fact, for what we have learned from the previous ensembles, for $N$ enough large we have
\begin{eqnarray}
\label{Rq1}
R^{(\mathrm{Q})}_{\mathsmaller{\Gamma}|\bm{h}}=\mathrm{O}\left(R^{(S)}_{\mathsmaller{\Gamma}}\right)=\mathrm{O}\left(R^{(A)}_{\mathsmaller{\Gamma}}\right).
\end{eqnarray} 
Eq. (\ref{Rq1}) can be derived either by using the fact that in the ensembles (S) (or (A)) and (B) we have and do not have large
fluctuations, respectively, or by using the fact that in the ensembles (S) (or (A)) and ($Q|_h$) we have and do not have large
fluctuations, respectively.

\section{Conclusions}
Real complex networks are the result of certain pseudo random processes that can be effectively described via hidden-variable models.
In this paper we have critically reviewed the definition of hidden-variable models and shown that, if the degree distribution $P(k)$
of the target real network is a power law, $P(k)\sim k^{-\gamma}$, the cut-off of the model must scale as $N^{\lambda}$ with $\lambda\geq 1$.
In fact, any choice with $\lambda<1$ leads to a difference between the scaling of the moments of $P(k)$ and the moments of the sampled degree $\mediaT{k^n}$ in the networks
generated from the hidden-variable model. We stress that this holds true for any $\gamma$.
We have then performed a detailed analysis of the relative fluctuations with the minimal/natural choice $\lambda=1$.
We have analyzed the behavior of the
relative fluctuations of the densities of the motifs, $n_\Gamma\propto\sum_i k_\Gamma(i)/N$, where $\Gamma$ represent any motif (Fig. \ref{fig1}),
and $k_\Gamma(i)$ is the generalized degree of node $i$, counting
how many motifs $\Gamma$ pass through it. We have then shown that, despite $n_\Gamma$ is an extensive observable,
if $\gamma\in[\gamma_1(\Gamma),\gamma_2(\Gamma)]$,
where $\gamma_1(\Gamma)\approx k_{\mathrm{min}}(\Gamma)$ and $\gamma_2(\Gamma)\approx 2 k_{\mathrm{max}}(\Gamma)$, $k_{\mathrm{min}}(\Gamma)$ and $k_{\mathrm{max}}(\Gamma)$ being the smallest and the largest degree of $\Gamma$,
$n_\Gamma$ is not-self-averaging, and a spin-glass picture is recovered (Eqs. (\ref{Png}) and (\ref{Png1})), with $R_\Gamma$ diverging for $N\to\infty$ (Figs. \ref{fig2}-\ref{fig7}).

We have shown that such a non-self-averaging behavior is only due to the variability of the expected degrees $\{h_i\}$,
whereas, when these are kept fixed, as happens when we are interested in reproducing only a given network realization with all possible graphs
having fixed expected degrees,
the fluctuations follow the standard law of large numbers $R_{\Gamma}\sim 1/\sqrt{N}$, like in classical random graphs.
Furthermore, we have seen that we recover the self-averaging limit (\ref{Png}) both when $\gamma\to\infty$,
and when $\gamma\to2$. Whereas the former is intuitively expected since $\gamma\to\infty$ corresponds, roughly speaking, to a an exponential
decay $P(k)\sim \exp(-bk)$, \textit{i.e.}, it is a classical random graph limit,
the latter is due to the fact that entropy of power-law random graphs goes to 0 when $\gamma\to2^+$ \cite{Bianconi3,Genio}.

If networks characterized by power law distributions are non self-averaging, network configurations, like in spin-glass models,
are intrinsically unpredictable. Furthermore, unless $\gamma$ is close to 2, or very large, the broad distribution \ref{Png1} makes a
power-law network effectively unstable to small perturbations. This in particular reflects on the stability/instability of communities \cite{Santo},
and more in general to the stability/instability of motifs on which the functionality of the network largely depends.

Moreover, we have seen that, even if we are dealing with a case of fixed expected degrees, 
the fact that our knowledge on the expected degree sequence $\{h_i\}$ is affected by unavoidable finite errors,
will result in a systematic bias between the simulations and the real network (Eq. (\ref{DD})).

The scenario we have described in this paper holds for hidden-variable models. Hidden-variable models describe equilibrium networks characterized
by soft-constraints (because only the expected degrees are fixed). Of course, in other classes of models fluctuations
might follow different rules. In particular networks can be described via models having hard constrains, or models built via a growing dynamics,
which may differ in many aspects from hidden-variable models. For example, in a model where the degrees are fixed, the degrees do not
fluctuate by construction. 
Nevertheless, we expect that the non-self-averaging or the spin-glass scenario seen in the hidden-variable models will have a counterpart in any model as soon as it is characterized
by power laws. 
We hope that our paper will stimulate further investigations in this directions.

\begin{acknowledgments}
This work was supported by DARPA grant No.\ HR0011-12-1-0012; NSF grants No.\ CNS-0964236 and CNS-1039646; and by Cisco Systems.
We thank D. Krioukov, K. Claffy and C. Orsini for useful discussions.
\end{acknowledgments}

\end{document}